\documentclass[preprints,article,accept,moreauthors,pdftex]{Definitions/mdpi}

\setitemize{parsep=6pt,itemsep=0pt,leftmargin=*,labelsep=5.5mm}
\setenumerate{parsep=6pt,itemsep=0pt,leftmargin=*,labelsep=5.5mm}
\setlist[description]{itemsep=0mm}

\usepackage{amsmath}%
\usepackage{amsfonts}%
\usepackage{amssymb}%
\usepackage{graphicx}

\usepackage[labelformat=simple]{subfig}

\graphicspath{{pics/}}

\usepackage{basic_math}
\usepackage{qm}
\usepackage{ifthen}
\def\bibfile{My_Library}% for central bibfile
\ifthenelse{\equal{\bibfile}{}}{%
}{%
}%

\usepackage[%
]{hyperref}

\firstpage{1}
\Title{Adaptive State Fidelity Estimation for Higher Dimensional Bipartite Entanglement}

\Author{Jun-Yi Wu %$^{1,\dagger,\ddagger}$\orcidA{},
}

\AuthorNames{Jun-Yi Wu,
}

\address [1]{%
Department of Physics, Graduate School of Science, The University of Tokyo, Hongo 7-3-1, Bunkyo-ku, Tokyo~113-0033, Japan; junyiwuphysics@gmail.com %This email is junyiwu@g.ecc.u-tokyo.ac.jp  or  junyiwuphysics@gmail.com?
\\
}

\abstract{
An adaptive method for quantum state fidelity estimation in bipartite higher dimensional systems is established. This method employs state verifier operators which are constructed by local POVM operators and adapted to the measurement statistics in the computational basis. Employing this method, the state verifier operators that stabilize Bell-type entangled states are constructed explicitly. Together with an error operator in the computational basis, one can estimate the lower and upper bounds on the state fidelity for Bell-type entangled states in few measurement configurations. These bounds can be tighter than the fidelity bounds derived in [Bavaresco et.al., Nature Physics (2018), 14, 1032–1037], if one constructs more than one local POVM measurements additional to the measurement in the computational basis.}


\begin{document}
%%%%%%%%%%%%%%%%%%%%%%%%%%%%%%%%%%%%%%%%%%

\section{Introduction}

Entanglement is the key resource in quantum information processing that brings advantages over its classical counterparts. In~many quantum information tasks, higher dimensional entanglement in qudit systems can fortify the power of quantum information processing over its applications in qubit systems, e.g., in~QKD~\cite{BrussMacchiavello2002-QutritQKD, CerfEtAlGisin2002-QKDDLvlSys}, quantum computation~\cite{NiuChuangShapiro2018-QditQCmp}, etc.
In practice, higher dimensional entangled states created in an entanglement generation process are always subjected to errors. To~qualify an entanglement generation process, one will need to extract some information on the created states by~measurements.

Employing quantum state tomography (QST), one can obtain the complete information of a quantum state~\cite{WoottersFields1989-StateDiscr,BanaszekEtAlSacchi1999-MLEQST, JamesEtAlWhite2001-MsmntQbt, ThewEtAlMunro2002-QditQST, FlammiaSilberfarbCaves2005-SICPOVMQST, GrossEtAlEisert2010-QSTCmprSns, AdamsonSteinberg2010-QSTMUB, MahlerEtAlSteinberg2013-AdptvQST, KalevHen2015-FdltyOptQST, PereiraEtAlDelgado2018-AdptvQSTQdit, StruchalinEtAlKulik2018-AdptvQST2Qdit}.
Although higher dimensional pure states can be determined using just five measurement settings~\cite{GoyenecheEtAlDelgado2015-5MsmntBsQST}, the~number of measurement configurations that are required in QST of a general $d$-dimensional quantum state scales badly with the dimension $d$.
For the qualification of a state generation process, instead of full QST, one may just need to employ quantum state fidelity estimation (QSFE) to reveal partial information about the most relevant Pauli operator components that signify the target state~\cite{GuehneEtAlPan2007-EntFidelityEst, WunderlichPlenio2009-EntFidelity, FlammiaLiu2011-FdltyEstPlMsmnt}.
One can even ease the measurement complexity, if~one just estimates the lower and upper bounds instead of the exact value of the state fidelity.
Such an approach is employed in~\cite{BavarescoEtAlHuber2018-2MUBsCrtfyHghDmEnt} for the detection of entanglement dimensionality in a higher-dimensional entanglement generation~process.

%Recently,
Another method for characterizing a quantum state resource called quantum state verification (QSV) is proposed in~\cite{PallisterLindenMontanaro2018-QSVLclMsmnt}. In~QSV, one tests a quantum state resource under eventual malicious attacks or errors by a quantum state verifier, which is also called a ``strategy''. One takes N samples from the inputs of the quantum state resource and verifies the samples by randomly selected local measurement setups assisted with classical communications. This method is generalized for noisy quantum state resources~\cite{YuShangGuhne2019-OptStVrf2P} and general adversary scenarios~\cite{ZhuHayashi2019-EffStVrfPRL, ZhuHayashi2019-EffStVrfPRA} with slightly different problem settings.
It is shown that the state verifier in QSV can also be exploited for state fidelity estimation, if~one continuously tests all the $N$ samples even when the testing fails~\cite{ZhuHayashi2019-EffStVrfPRA}.
The frequency of passing a test in the $N$ samples will then determine a lower and an upper bound on the state fidelity with certain confidence~levels.

Since lower bounds on quantum state fidelity can be employed to detect the entanglement dimensionality of a bipartite state~\cite{BavarescoEtAlHuber2018-2MUBsCrtfyHghDmEnt}, a~tighter lower bound on quantum state fidelity means better robustness of the entanglement detection against noises in a system.
In both QSFE and QSV, for~each copy of a testing state, one randomly chooses a measurement setting from a set of predefined measurement configurations to obtain the statistics regardless of the particular noises in an individual entanglement generation.
The bounds on state fidelity obtained in such predefined measurement configurations are in general not optimum under these particular noises.
In practice, the feasibility and efficiency of different local measurement configurations differ from each other. Some measurement configurations, e.g.,~measurements in the computational basis, are much easier and more efficient to implement than the other configurations, e.g.,~POVM measurements in a non-computational basis.
Instead of randomly choosing a measurement setting from predefined measurement configurations for each copy of a testing state, it is possible to efficiently obtain the information of diagonal elements of a quantum state density matrix in the computational basis prior to the other measurement settings.
This~information contains some partial information about the noises in a generation process of a bipartite quantum state.
One can therefore exploit this information to tailor the subsequential measurement settings for the particular noises of the testing system to refine the state fidelity estimation, which would be important for entanglement detection subject to noises.

In this paper, we will employ the state verifiers, which are introduced in QSV, to~derive the lower and upper bounds on state fidelity of bipartite qudit states for the purpose of QSFE.
We will show in Lemma \ref{lemma::SF_bounds} that measurement statistics in the computational basis can be exploited to refine the bounds on quantum state fidelity derived from state verifiers.
Since these refined bounds depend both on the measurement statistics in the computational basis $\mathbb{P}_{e}$ and the configurations of subsequential measurements $\mathbb{M}$, one can adapt the subsequential measurement configurations $\mathbb{M}$ for tighter bounds on state fidelity subject to the a priori statistics $\mathbb{P}_{e}$.
Following this idea, we will derive an adaptive state fidelity estimation approach for bipartite Bell-type states in Theorem \ref{theorem::fidelity_bounds}.
We will compare our approach with the one derived in~\cite{BavarescoEtAlHuber2018-2MUBsCrtfyHghDmEnt} and demonstrate it under different types of~noises.

\section{Results}

%%%%%%%%%%%%%%%%%%%%%%%%%%%%%%%%%%%%%%%%%%
\subsection{Quantum State Fidelity Estimation Employing State~Verifiers}

In a quantum information processing employing a pure state $\ket{\psi}$ in a bipartite $d$-dimensional system $\mathbb{H}_{d}^{(A)}\otimes\mathbb{H}_{d}^{(B)}$, the~very first task is to create bipartite quantum states as close as possible to the target state $\ket{\psi}$.
To evaluate how good a state preparation is, one can estimate the $\ket{\psi}$-state fidelity $F_{\psi}$ of the generated states $\widehat{\rho}$ in local measurements, where the state fidelity $F_{\psi}$ is defined as
\begin{equation}
  F_{\psi}(\rho) := \braket{\psi|\widehat{\rho}|\psi}.
\end{equation}

In this section, we review the strategy operators employed in QSV~\cite{PallisterLindenMontanaro2018-QSVLclMsmnt, YuShangGuhne2019-OptStVrf2P, ZhuHayashi2019-EffStVrfPRL, ZhuHayashi2019-EffStVrfPRA, LiHanZhu2019-EffStVrf2P}, and~their application in QSFE.
In QSFE, one evaluates expectation values of certain observables from the whole measurement outputs instead of testing each input by each output of measurements according to a ``strategy''; we~therefore refer to the ``strategy'' in QSV as ``state verifier operators'' in the context of QSFE in this~paper.

In the measurement of the computational basis $\{\ket{e_{k_{A}},e_{k_{B}}}\}_{k_{A},k_{B}}$, one can verify the testing state $\widehat{\rho}$ by the characteristic correlations of the target state $\ket{\psi}$.
The probability of the outputs satisfying the target characteristic correlations is determined to the expectation value of the following $\ket{\psi}$-state stabilizer:
\begin{equation}
  \widehat{V}_{e} := \sum_{k_{A},k_{B}: \braket{e_{k_{A}},e_{k_{B}}|\psi}\neq0} \projector{e_{k_{A}},e_{k_{B}}}.
\end{equation}

We call a stabilizer of the target state $\ket{\psi}$ a $\ket{\psi}$-state verifier.
If the measurement in the Schmidt basis of $\ket{\psi}$ is feasible and efficient in a laboratory, it is preferable to choose the Schmidt basis as the computational basis, since the state verifier $\widehat{V}_{e}$ constructed in the Schmidt basis has the least rank, which means that $\widehat{V}_{e}$ can detect the $\ket{\psi}$-orthogonal part of a testing state $\widehat{\rho}$ more efficiently.

To estimate the quantum state fidelity, a~single state verifier in the computational basis is not enough, since $\ket{\psi}$ is not the only one state that is stabilized by $\widehat{V}_{e}$.
To construct a state verifier that stabilizes only the target state $\ket{\psi}$, one needs to include the state verifiers in the other measurement basis.
Let $\mathbb{M}$ a set of measurement configurations additional to the computational basis
\begin{equation}
\label{eq::msmnt_config}
  \mathbb{M} := \{\mathcal{M}_{j}^{(A)}\otimes\mathcal{M}_{j}^{(B)}\}_{j}
  \;\;\text{ with }\;\;
  \mathcal{M}_{j}
  :=
  \{\widehat{M}_{m}(j)\}_{m=0, ..., d},
\end{equation}
where $\mathcal{M}_{j}^{(A,B)}$ are POVM measurements in the $d$-dimensional local system $\mathbb{H}_{d}^{(A,B)}$.
The POVM measurements $\mathcal{M}_{j}^{(A,B)}$ are constructed with $d+1$ measurement operators $\widehat{M}_{m}(j)$, which are projections onto the corresponding measurement-basis states $\{\ket{E_{m}(j)}\}_{m}$,
\begin{equation}
\label{eq::msmnt_op}
  \widehat{M}_{m}(j)
  :=
  \left\{
    \begin{array}{ll}
      \frac{1}{d}\projector{E_{m}(j)}, & m=0, ..., d-1; \\
      \id - \sum_{m=0}^{d-1}\widehat{M}_{m}, & m=d.
    \end{array}
  \right.
\end{equation}

Note that, for projective measurements with orthogonal basis states, there is no need to add the factor $1/d$ in Equation \eqref{eq::msmnt_op}.
However, for~consistency of formulation, we adopt the representation in Equation \eqref{eq::msmnt_op} for projective measurements.
In each measurement configuration $\mathcal{M}_{j}^{(A)}\otimes\mathcal{M}_{j}^{(B)}$, one~can~construct a state verifier operator $\widehat{V}_{j}$ by adding up its corresponding measurement operators $\widehat{M}_{m_{A}}(j)\otimes\widehat{M}_{m_{B}}(j)$ with weights $v_{m_{A}m_{B}}$, such that $\widehat{V}_{j}$ stabilizes the target state.
\begin{Lemma}[Construction of a state verifier in local POVM measurements]
\label{lemma::Vj_contruction}
  The state verifier $\widehat{V}_{j}$ in the measurement configuration $\mathcal{M}_{j}^{(A)}\otimes\mathcal{M}_{j}^{(B)}$ that stabilizes $\ket{\psi}$ can be explicitly constructed by
\begin{equation}
  \label{eq::Vj_constr}
    \widehat{V}_{j} :=
    \sum_{m_{A},m_{B}=0}^{d-1} v_{m_{A}m_{B}}(j)
    \;
    \widehat{M}_{m_{A}}(j)\otimes\widehat{M}_{m_{B}}(j).
  \end{equation}%

  Here, the~weights $v_{m_{A}m_{B}}$ are determined by a transformation operator \textls[-15]{$\widehat{T}_{A,B}(j):=\sum_{m=0}^{d-1}\ket{E_{m}^{(A,B)}(j)}\bra{e_{m}^{(A,B)}}$} that maps the local computational basis states $\{\ket{e_{m}^{(A,B)}}\}_{m}$ to the measurement basis~states $\{\ket{E_{m}^{(A,B)}(j)}\}_{m}$ associated with the local POVM $\mathcal{M}_{j}^{(A,B)}$ as follows:
\begin{equation}
  \label{eq::Vj_constr_vmAmB}
    v_{m_{A}m_{B}}
    =
    \left\{
      \begin{array}{ll}
        d^{2}\frac{
          \braket{e_{m_{A}},e_{m_{B}}|\widehat{T}_{A}^{-1}(j)\otimes \widehat{T}_{B}^{-1}(j)|\psi}
          %\braket{\psi|\widehat{T}_{A}^{-1}(j)\otimes\widehat{T}_{B}^{-1}(j)|e_{m_{A}},e_{m_{B}}}
        }{
          \braket{e_{m_{A}},e_{m_{B}}|\widehat{T}_{A}^{\dagger}(j)\otimes \widehat{T}_{B}^{\dagger}(j)|\psi}
        }
        & ,\text{ for }
        \braket{e_{m_{A}},e_{m_{B}}|\widehat{T}_{A}^{\dagger}(j)\otimes \widehat{T}_{B}^{\dagger}(j)|\psi}\neq0
        \\
        0
        & ,\text{ for }
        \braket{e_{m_{A}},e_{m_{B}}|\widehat{T}_{A}^{\dagger}(j)\otimes \widehat{T}_{B}^{\dagger}(j)|\psi}=0
      \end{array}
    \right. .
  \end{equation}

\end{Lemma}

\begin{proof} see \textbf{Methods}.%Please clearly proof where it ends.
\end{proof}
A good measurement configuration $\mathcal{M}_{j}^{(A)}\otimes\mathcal{M}_{j}^{(B)}$ should have nonzero $v_{m_{A}m_{B}}$ in its state verifier $\widehat{V}_{j}$ as few as possible, which leads to the minimum rank of $\widehat{V}_{j}$ and better detection efficiency of $\ket{\psi}$-orthogonal states.
For this reason, POVM measurements are preferable for most bipartite states in general. For~example, for~the general Bell-type states that will be studied in Section~\ref{sec::QSFE_BType}, the~POVM measurements that are associated with the generalized Heisenberg--Weyl operators defined in Equation~\eqref{eq::pl_op} lead to the state verifiers derived in Equation \eqref{eq::st_verifier_Btype}, which have the minimum rank of $d$.
For the maximally entangled states, the~projective measurements in the mutually unbiased bases are the optimum configurations.
In this case, the~state verifiers $\{\widehat{V}_{j}\}_{j\in\mathbb{M}}$ are local unitary transformations of $\widehat{V}_{e}$.

By mixing the state verifiers $\{\widehat{V}_{j}\}_{j\in\mathbb{M}}$ that are associated with the measurement settings in $\mathbb{M}$, one~can construct a state verifier $\widehat{V}_{\mathbb{M}}$,
\begin{equation}
\label{eq::psi_verifier_M}
  \widehat{V}_{\mathbb{M}}
  :=
  \sum_{j\in\mathbb{M}}u_{j}\widehat{V}_{j}
  \;\;\text{ with }\;\;
  \sum_{j\in\mathbb{M}} u_{j} = 1.
\end{equation}
Together with the state verifier $\widehat{V}_{e}$ in the computational basis, one can then construct a $\ket{\psi}$-state verifier operator, which only stabilizes the target state $\ket{\psi}$,
\begin{equation}
\label{eq::psi_verifier_def}
  \widehat{V}_{\psi} :=
  u_{e}\widehat{V}_{e} + (1-u_{e})\widehat{V}_{\mathbb{M}}
  \;\;\text{ with }\;\;
  0\le u_{e}\le 1.
\end{equation}
Since the $\ket{\psi}$-state verifier $\widehat{V}_{\psi}$ is a Hermitian stabilizer of $\ket{\psi}$ by definition, the~$\ket{\psi}$-state verifier can be decomposed into the mixture of the projection onto the target state and its orthogonal part $\widehat{V}_{\psi}^{\bot}$, i.e.,~$\widehat{V}_{\psi}=\projector{\psi} +  \widehat{V}_{\psi}^{\bot}$ with $\braket{\psi|\widehat{V}_{\psi}^{\bot}|\psi} = 0$.
Note that the state verifier $\widehat{V}_{\psi}$ is called a verification strategy in the context of quantum state verification (QSV).
Let $\{\lambda_{i}\}_{i}$ be the eigenvalues of the $\ket{\psi}$-orthogonal operator $\widehat{V}_{\psi}^{\bot}$ associated with the eigenstates $\{\ket{\phi_{i}}\}_{i}$.
The maximum and minimum eigenvalue $\lambda_{\max,\min}$ of $\widehat{V}_{\psi}^{\bot}$ determines the efficiency of the verification strategy in QSV as well as the fidelity bounds in QSFE~\cite{ZhuHayashi2019-EffStVrfPRA},
\begin{equation}
\label{eq::SF_bounds_nonadaptive}
  \frac{\braket{\widehat{V}_{\psi}} - \lambda_{\max}}{1 - \lambda_{\max}}
  \le F_{\psi} \le
  \frac{\braket{\widehat{V}_{\psi}} - \lambda_{\min}}{1 - \lambda_{\min}}.
\end{equation}

\noindent Let $\ket{\phi_{\max}}$ and $\ket{\phi_{\min}}$ be the eigenstates of $\widehat{V}_{\psi}^{\bot}$ associated with the maximum and minimum eigenvalues $\lambda_{\max}$ and $\lambda_{\min}$, respectively.
The lower bound in Equation \eqref{eq::SF_bounds_nonadaptive} can be achieved by the testing states $\widehat{\rho}\in \spn\left(\ket{\psi},\ket{\phi_{\max}}\right)$ in the Hilbert subspace that is spanned by the target state $\ket{\psi}$ and the maximum-eigenvalue state $\ket{\phi_{\max}}$, while the upper bound can be achieved by the states $\widehat{\rho}\in\spn\left(\ket{\psi},\ket{\phi_{\min}}\right)$.
However, the~noises in a state generation process are in general not the eigenstates $\ket{\phi_{\max}}$ or $\ket{\phi_{\min}}$ of the operator $\widehat{V}_{\psi}^{\bot}$, which means that the bounds in Equation \eqref{eq::SF_bounds_nonadaptive} are not the tightest for a particular noisy state generation.
As the fidelity lower bound can be employed for entanglement dimensionality certification~\cite{BavarescoEtAlHuber2018-2MUBsCrtfyHghDmEnt}, a~tighter fidelity lower bound in a state fidelity estimation implies the better robustness of the entanglement detection against the noises that present in the experiment.
It is therefore desirable to refine the fidelity bounds in QSFE by adapting the estimation approach to the noises of a particular state generation.

%\bigskip

\subsection{Quantum State Fidelity Estimation Assisted with Measurement Statistics in the Computational Basis}

In this section, we employ the state verifiers in a scenario of quantum state fidelity estimation under the assumption that the computational-basis measurement is more efficient and feasible than the other measurement configurations.
In this case, one can first measure a testing state $\widehat{\rho}$ in the computational basis and obtain the corresponding measurement statistics:
\begin{equation}
\label{eq::prob_in_compbasis}
  \mathbb{P}_{e}
  :=
  \{\prob_{e} (k_{A},k_{B})\}_{k_{A},k_{B}}
  =
  \{\braket{e_{k_{A}},e_{k_{B}} | \widehat{\rho}| e_{k_{A}},e_{k_{B}}}\}_{k_{A},k_{B}}.
\end{equation}
This measurement statistics contains information about the noises in a state generation.
These noises can contribute to the expectation value of the $\ket{\psi}$-orthogonal part $\widehat{V}_{\psi}^{\bot}$ of the state verifier $\widehat{V}_{\psi}$,
\begin{equation}
  \widehat{V}_{\psi}^{\bot}
  =
  u_{e}\widehat{V}_{e}^{\bot}+(1-u_{e})\widehat{V}_{\mathbb{M}}^{\bot},
\end{equation}
where $\widehat{V}_{e}^{\bot}$ and $\widehat{V}_{\mathbb{M}}^{\bot}$ are the $\ket{\psi}$-orthogonal part of $\widehat{V}_{e}$ and $\widehat{V}_{\mathbb{M}}$, respectively,
\begin{equation}
  \widehat{V}_{e}^{\bot} = \widehat{V}_{e} - \projector{\psi}
  \;\;\text{ and }\;\;
  \widehat{V}_{\mathbb{M}}^{\bot} = \widehat{V}_{\mathbb{M}} - \projector{\psi}.
\end{equation}
To estimate the state fidelity $F_{\psi}$, one will need to exclude the contribution of $\ket{\psi}$-orthogonal part $\braket{\widehat{V}_{\psi}^{\bot}}$ from the expectation value of the state verifier $\braket{\widehat{V}_{\psi}}$, since $F_{\psi} = \braket{\widehat{V}_{\psi}} - \braket{\widehat{V}_{\psi}^{\bot}}$.
In Equation \eqref{eq::SF_bounds_nonadaptive}, the~expectation value $\braket{\widehat{V}_{\psi}^{\bot}}$ is bounded by its maximum and minimum eigenvalues,
\begin{equation}
\label{eq::Vpsi_bot_bounds_lambda}
  \lambda_{\max}(1-F_{\psi})\ge\braket{\widehat{V}_{\psi}^{\bot}}\ge\lambda_{\min}(1-F_{\psi}),
\end{equation}
which does not depend on the measurement statistics $\mathbb{P}_{e}$.
Here, the~a priori information of the computational-basis measurement statistics $\mathbb{P}_{e}$ can help us to adjust the measurement configurations $\mathbb{M}$ to the noises of the systems and refine the bounds on the expectation value $\braket{\widehat{V}_{\psi}^{\bot}}$.

To estimate $\braket{\widehat{V}_{\psi}^{\bot}}$ exploiting the measurement statistics $\mathbb{P}_{e}$, one can bound the operator $\widehat{V}_{\psi}^{\bot}$ by an operator $\widehat{\mathcal{I}}$, which is diagonal in the computational basis,
\begin{equation}
\label{eq::Phi_comp_basis}
  \widehat{\mathcal{I}}
  =
  \widehat{V}_{e} + \widehat{\mathcal{E}}_{\mathbb{M}},
\end{equation}
where $\widehat{\mathcal{E}}_{\mathbb{M}}$ is the non-zero diagonal part of the $\ket{\psi}$-orthogonal operator $\widehat{V}_{\mathbb{M}}^{\bot}$ assigned by a weight $\gamma_{k_{A}k_{B}}$,
\begin{equation}
\label{eq::err_op}
  \widehat{\mathcal{E}}_{\mathbb{M}}
  =
  \sum_{k_{A},k_{B}: \braket{e_{k_{A}},e_{k_{B}}|\widehat{V}_{\mathbb{M}}^{\bot}|e_{k_{A}},e_{k_{B}}}\neq 0}\gamma_{k_{A}k_{B}}
  \projector{e_{k_{A}},e_{k_{B}}}.
\end{equation}
The operator $\widehat{\mathcal{E}}_{\mathbb{M}}$ contains the information of the $\ket{\psi}$-orthogonal contributions in $\widehat{V}_{\mathbb{M}}$, which are the errors that we want to exclude from the state verifier.
This information can be extracted from the measurement statistics $\mathbb{P}_{e}$ in the computational basis by the operator $\widehat{\mathcal{I}}$ prior to the implementation of the measurement $\mathbb{M}$.
It can help us to evaluate the measurement configurations $\mathbb{M}$ and to bound the operator $\widehat{V}_{\psi}^{\bot}$ exploiting the a priori statistics $\mathbb{P}_{e}$.
The operator $\widehat{\mathcal{I}}$ can be decomposed into the $\ket{\psi}$ projector and a non-$\ket{\psi}$ component $\widehat{\mathcal{I}}^{\bot}$,
\begin{equation}
  \widehat{\mathcal{I}} =
%  \widehat{V}_{e} + \widehat{\mathcal{E}}_{\mathbb{M}}(\mathbb{P}_{e})
  \projector{\psi} +
  \widehat{\mathcal{I}}^{\bot}
  \;\;\text{ with }\;\;
  \widehat{\mathcal{I}}^{\bot}
  =
  \widehat{V}_{e}^{\bot} +
  \widehat{\mathcal{E}}_{\mathbb{M}}.
\end{equation}
The expectation value $\braket{\widehat{\mathcal{I}}}$ is the sum of the state fidelity $F_{\psi}$ and the expectation value $\braket{\widehat{\mathcal{I}}^{\bot}}$, which~contains partial information about the $\ket{\psi}$-orthogonal contribution $\braket{\widehat{V}_{\psi}^{\bot}}$ of a testing state in the expectation value of the state verifier $\braket{\widehat{V}_{\psi}}$.
One can show that there exists an assignment of the weights $\gamma_{k_{A}k_{B}}$ in $\widehat{\mathcal{E}}_{\mathbb{M}}$, such that the operators $\widehat{V}_{\psi}^{\bot}$ and $\widehat{\mathcal{I}}^{\bot}$ can be decomposed by a set of pure state $\{\ket{\widetilde{\phi}_{i}}\}_{i}$,
\begin{equation}
\label{eq::psi_Phi_decomp}
  \widehat{V}_{\psi}^{\bot}
  =
  \sum_{i} \tilde{\lambda}_{i}\projector{\widetilde{\phi}_{i}}
%  \;\;\text{ with }\;\;
% \widetilde{\lambda}_{i}\ge0
  \;\;\text{ and }\;\;
%\label{eq::psi_Phi_decomp}
  \widehat{\mathcal{I}}^{\bot}
  =
  \sum_{i}r_{i}\projector{\widetilde{\phi}_{i}}
%  \;\;\text{ with }\;\;
%  r_{i}>0,
\end{equation}
where $\{\ket{\widetilde{\phi}_{i}}\}_{i}$ are in general non-orthogonal, $\widetilde{\lambda}_{i}\ge0$ are non-negative and $r_{i}>0$ are positive.
One can then bound the operator $\widehat{V}_{\psi}^{\bot}$ by $\widehat{\mathcal{I}}^{\bot}$ with two real-value coefficients $\alpha$ and $\beta$ such that $\alpha\widehat{\mathcal{I}}^{\bot}\succeq\widehat{V}_{\psi}^{\bot}\succeq\beta\widehat{\mathcal{I}}^{\bot}$,
which refines the bounds on the $\ket{\psi}$-orthogonal contribution $\braket{\widehat{V}_{\psi}^{\bot}}$ in $\braket{\widehat{V}_{\psi}}$ given in Equation \eqref{eq::Vpsi_bot_bounds_lambda},
\begin{equation}
\label{eq::V_psi_bot_bound}
  \alpha\left(\braket{\widehat{\mathcal{I}}} - F_{\psi}\right)
  \ge
  \braket{\widehat{V}_{\psi}^{\bot}}
  \ge
  \beta\left(\braket{\widehat{\mathcal{I}}} - F_{\psi}\right).
\end{equation}
As a result, one can then refine the bounds on the state fidelity given in Equation \eqref{eq::SF_bounds_nonadaptive} as follows.

\begin{Lemma}[Bounds on state fidelity]
\label{lemma::SF_bounds}
  The state fidelity for a target state $\ket{\psi}$ is bounded by
\begin{equation}
  \label{eq::SF_bounds}
    \frac{
      \braket{\widehat{V}_{\psi}} - \alpha\braket{\widehat{\mathcal{I}}}
    }{
      1-\alpha
    }
    \le F_{\psi} \le
    \frac{
      \braket{\widehat{V}_{\psi}} - \beta\braket{\widehat{\mathcal{I}}}
    }{
      1-\beta
    },
  \end{equation}
  where $\alpha$ and $\beta$ are the maximum and minimum ratio between $\tilde{\lambda}_{i}$ and $r_{i}$
\begin{equation}
  \label{eq::alpha_beta}
    \alpha := \max_{i} \frac{\tilde{\lambda}_{i}}{r_{i}}
    \;\;\text{ and }\;\;
    \beta := \min_{i} \frac{\tilde{\lambda}_{i}}{r_{i}}.
  \end{equation}
% of the ratios between $\tilde{\lambda}_{i}$ and $r_{i}$,
%  \begin{equation}
%    \alpha := \max \frac{\tilde{\lambda}_{i}}{r_{i}}
%    \;\;\text{ and }\;\;
%    \beta := \min \frac{\tilde{\lambda}_{i}}{r_{i}}
%  \end{equation}
%\noindent Proof: see \textbf{Methods}.
\end{Lemma}
\begin{proof} see \textbf{Methods}.
\end{proof}
\noindent
A trivial construction of $\widehat{\mathcal{E}}_{\mathbb{M}}$ is the assignment of $\gamma_{k_{A}k_{B}}=1$, which leads to $\widehat{\mathcal{I}}=\widehat{\id}$.
For this construction, the~decomposition in Equation \eqref{eq::psi_Phi_decomp} is the eigenstate decomposition of $\widehat{V}_{\psi}^{\bot}$.
In this case, the~bounds in Equation \eqref{eq::SF_bounds} coincide with the bounds given in Equation \eqref{eq::SF_bounds_nonadaptive}.
Since $\braket{\widehat{\id}}=1$ is constant and does not depend on the measurement configurations $\mathbb{M}$ and measurement statistics $\mathbb{P}_{e}$ in the computational basis, it can not be employed to adapt the measurement configurations $\mathbb{M}$ to $\mathbb{P}_{e}$.

In order to adapt the measurement configurations $\mathbb{M}$ to $\mathbb{P}_{e}$, one needs to introduce the $\mathbb{M}$ and $\mathbb{P}_{e}$ dependency in $\braket{\widehat{\mathcal{I}}}$, such that one can find the optimal measurement configuration $\mathbb{M}$ for the minimum $\braket{\widehat{\mathcal{I}}}$ subject to a given measurement statistics $\mathbb{P}_{e}$.
To this end, one can explicitly construct a nontrivial $\widehat{\mathcal{I}}$ and determine the coefficients $(\alpha,\beta)$ following the protocol given in the proof of Lemma \ref{lemma::SF_bounds} in Section~\ref{sec::methods} (Methods).
Employing the operator $\widehat{\mathcal{I}}$ constructed in Equation \eqref{eq::Phi_construction_Z_class}, one can then adapt the measurement configurations $\mathbb{M}$ to the measurement statistics $\mathbb{P}_{e}$ such that the expectation value $\braket{\widehat{\mathcal{I}}}$ is minimum subject to a given $\mathbb{P}_{e}$, which leads to a higher lower bound on the state fidelity.
Usually, the~coefficient $\beta$ is zero, unless~one chooses a large set of measurement configurations such that the state verifier $\widehat{V}_{\psi}$ has the same rank as $\widehat{\mathcal{I}}$.
As a consequence, the~minimization of $\braket{\widehat{\mathcal{I}}}$ does not affect the upper bound in most cases.
Following these steps, one can therefore construct the subsequential measurements $\mathbb{M}$ depending on the measurement statistics in the computational basis $\mathbb{P}_{e}$, which means the operators $\widehat{V}_{\psi}$ and $\widehat{\mathcal{I}}$ in Equation \eqref{eq::SF_bounds} also depend on $\mathbb{P}_{e}$,
\begin{equation}
  \widehat{V}_{\psi} = \widehat{V}_{\psi}(\mathbb{P}_{e})
  \;\;\text{ and }\;\;
  \widehat{\mathcal{I}} = \widehat{\mathcal{I}}(\mathbb{P}_{e}).
\end{equation}
As a result, Lemma \ref{lemma::SF_bounds} allows us to estimate quantum state fidelity employing $\widehat{V}_{\psi}(\mathbb{P}_{e})$ and $\widehat{\mathcal{I}}(\mathbb{P}_{e})$ adapted to the measurement statistics in the computational basis $\mathbb{P}_{e}$ to obtain tighter bounds.
In the next section, we will employ this method to derive an adaptive approach of quantum state fidelity estimation for Bell-type states explicitly.

\subsection{Adaptive~State Fidelity Estimation for Bell-Type States}
\label{sec::QSFE_BType}

A general Bell-type entangled state in $d\times d$ Hilbert state is an entangled state with the Schmidt rank $d$, which is an important higher dimensional entanglement resource in bipartite systems.
If the Schmidt basis happens to be more feasible than the other basis in a laboratory, one can employ the Schmidt basis as the computational basis in our adaptive estimation approach.
In this case, a~bipartite pure state is decomposed as
\begin{equation}
\label{eq::Schmidt_decomp}
  \ket{\psi} = \sum_{k=0}^{d-1} s_{k}\ket{e_{k}^{(A)},e_{k}^{(B)}}
  \;\;\text{ with }\;\;
  s_{k}>0,
\end{equation}
where $s_{k}$ are the Schmidt coefficients.
In order to construct a state verifier for a Bell-type state $\ket{\psi}$, one needs to construct stabilizers of $\ket{\psi}$ employing measurement operators in different measurement bases.
In the computational basis, the~state verifier $\widehat{V}_{e}$ that characterizes the correlations of the target state $\ket{\psi}$ is given by
\begin{equation}
  \widehat{V}_{e} = \sum_{k}\projector{e_{k},e_{k}}.
\end{equation}

For the construction of state verifiers in the other measurement bases, one needs the other stabilizers of the Bell-type state $\ket{\psi}$, which can be derived from the standard Heisenberg--Weyl (HW) operators~\cite{DurtAtElZyczkowski2010-MUBs} with a modification associated with a coefficient vector $\vec{\chi} = (\chi_{0}, ..., \chi_{d-1})$.
A $\vec{\chi}$-modified HW operator $\widehat{\Omega}_{i,j}(\vec{\chi})$ is comprised of the $\vec{\chi}$-modified shift operator $\widehat{X}(\vec{\chi})$ and the clock operator $\widehat{Z}$,
\begin{equation}
\label{eq::pl_op}
  \widehat{\Omega}_{i,j}(\vec{\chi}) := w^{-\frac{ij}{2}(d-1)}\widehat{X}^{i}(\vec{\chi})\widehat{Z}^{j},
\end{equation}
where the $\vec{\chi}$-modified shift operator $\widehat{X}(\vec{\chi})$ and the clock operator $\widehat{Z}$ are defined as
\begin{gather}
  \widehat{X}(\vec{\chi}): = \sum_{k=0}^{d-1} \frac{\chi_{k\oplus1}}{\chi_{k}}\ket{e_{k\oplus 1}}\bra{e_{k}}
  \;\;\text{ and }\;\;
  \widehat{Z}: = \sum_{k=0}^{d-1}w^{k}\projector{e_{k}}
\end{gather}
with $|\vec{\chi}| = 1$ and $w := e^{\imI \frac{2\pi}{d}}$.
Here, the~symbol ``$\oplus$'' stands for the $d$-modulus plus \footnote{The symbol $\oplus_{d}$ ($\ominus_{d}$) is employed to denote the $d$-modulus plus (minus) of two quantities, e.g.,~$a \oplus_{d} b:=(a+b)_{\pmod d}$ and $a \ominus_{d} b:=(a-b)_{\pmod d}$. For~conciseness, we omit the subscript $d$.}.
Note that the relevant HW operators in this paper are the operators with the label $i=1$, of~which the notation are simplified by $\widehat{\Omega}_{j}:=\widehat{\Omega}_{1,j}$.
The target Bell-type state $\ket{\psi}$ is stabilized by all the local HW operators $\{\widehat{\Omega}_{j}(\vec{\chi}_{A})\otimes\widehat{\Omega}_{-j}(\vec{\chi}_{B})\}_{j=0,...,d-1}$ with the modification coefficients $\vec{\chi}_{A,B}$ satisfying
\begin{equation}
\label{eq::chi_coeff_and_sCoeff}
  s_{k} = \frac{\chi^{(A)}_{k}\chi^{(B)}_{k}}{\sqrt{\sum_{k}|\chi^{(A)}_{k}\chi^{(B)}_{k}|^{2}}}
  \;\;\text{ for all }k.
\end{equation}
As a consequence, the~measurement configurations $\mathbb{M}$ for the $\ket{\psi}$-state verifier can be constructed in the eigenbasis of the $\vec{\chi}_{A,B}$-modified HW operators,
%selected from the set of the following local measurements
\begin{equation}
\label{eq::ms_config}
  \mathbb{M}(\vec{\chi}_{A},\vec{\chi}_{B}) \subseteq
  \{
    \mathcal{M}[\widehat{\Omega}_{j}(\vec{\chi}_{A})]
    \otimes
    \mathcal{M}[\widehat{\Omega}_{-j}(\vec{\chi}_{B})]
    :j=0, ..., d-1
  \}.
%  \{\mathcal{M}_{j}(\vec{\chi}_{A}(\mathbb{P}_{e}))\otimes\mathcal{M}_{-j}(\vec{\chi}_{B}(\mathbb{P}_{e})): j=0, ..., d-1\},
\end{equation}
where the local POVM measurement $\mathcal{M}[\widehat{\Omega}_{j}(\vec{\chi})]
= \{\widehat{M}_{m}[\widehat{\Omega}_{j}(\vec{\chi})]\}_{m = 0, ..., d}$ in the $\widehat{\Omega}_{j}(\vec{\chi})$ eigenbasis $\{\ket{E_{m}(j;\vec{\chi})}\}_{m}$   consists of the measurement operators $\widehat{M}_{m}[\widehat{\Omega}_{j}(\vec{\chi})] = \projector{E_{m}(j; \vec{\chi})}/d$ as defined in Equation \eqref{eq::msmnt_op}.
To implement such a measurement, one has to know the explicit form of the $\widehat{\Omega}_{j}$ eigenstates $\{\ket{E_{m}(j;\vec{\chi})}\}_{m}$ in the computational basis, which are constructed by
\begin{equation}
\label{eq::HW_op_basis}
  \ket{E_{m}(j; \vec{\chi})}
  :=
  \sum_{k=0}^{d-1}
  w^{-(m+\frac{1}{2}jd)k+\frac{1}{2}jk^{2}}\chi_{k}\ket{e_{k}}.
\end{equation}
As one can show that $\widehat{\Omega}_{j}\ket{E_{m}(j;\vec{\chi})} = w^{m}\ket{E_{m}(j;\vec{\chi})}$ by simply applying $\widehat{\Omega}_{j}$ on the state, the~eigenstate $\ket{E_{m}(j;\vec{\chi})}$ is associated with the eigenvalue $w^{m}$.
Since the eigenstate $\ket{E_{m}(j;\vec{\chi})}$ depends on the coefficient $\vec{\chi}$,
the set of measurement configurations $\mathbb{M}$ are therefore determined by the coefficients $\vec{\chi}_{A,B}$, which can be adapted to the measurement statistics $\mathbb{P}_{e}$ in the computational basis, i.e.,~$\vec{\chi}_{A,B} = \vec{\chi}_{A,B}(\mathbb{P}_{e})$.
In each measurement configuration $\mathcal{M}[\widehat{\Omega}_{j}(\vec{\chi}_{A})]\otimes\mathcal{M}[\widehat{\Omega}_{-j}(\vec{\chi}_{B})]$,
one can construct its corresponding state verifier $\widehat{V}_{j}$ according to Lemma \ref{lemma::Vj_contruction},
\begin{equation}
\label{eq::st_verifier_Btype}
  \widehat{V}_{j}(\vec{\chi}_{A}, \vec{\chi}_{B})
  =
  \frac{d}{ \sum_{k}|\chi_{k}^{(A)}\chi_{k}^{(B)}|^{2} }
  \sum_{m=0}^{d-1}
  \widehat{M}_{m}[\widehat{\Omega}_{j}(\vec{\chi}_{A})]
  \otimes
  \widehat{M}_{-m}[\widehat{\Omega}_{-j}(\vec{\chi}_{B})].
\end{equation}
The state verifier $\widehat{V}_{j}$ has the minimum rank of $d$, which is optimum for a Bell-type state $\ket{\psi}$ in a $d\times d$-dimensional Hilbert space.
The state verifier $\widehat{V}_{\mathbb{M}}=\sum_{j}u_{j}\widehat{V}_{j}$ associated with the non-computational-basis measurement configurations $\mathbb{M}$ is then comprised of $\{\widehat{V}_{j}\}_{j\in\mathbb{M}}$ with certain weights $\{u_{j}\}_{j\in\mathbb{M}}$ according to Equation \eqref{eq::psi_verifier_M}.

Together with the state verifier $\widehat{V}_{e}$ in the computational basis, one can construct a $\ket{\psi}$-state verifier $\widehat{V}_{\psi}=u_{e}\widehat{V}_{e}+(1-u_{e})\widehat{V}_{\mathbb{M}}$ according to Equation \eqref{eq::psi_verifier_def}.
To estimate the state fidelity, one still needs to construct the operator $\widehat{\mathcal{I}} = \widehat{V}_{e}+\widehat{\mathcal{E}}_{\mathbb{M}}$, where the error operator $\widehat{\mathcal{E}}_{\mathbb{M}}$ can be determined according to Equation \eqref{eq::Phi_construction_Z_class} as follows:
\begin{equation}
\label{eq::all_Bell_op}
  \widehat{\mathcal{E}}_{\mathbb{M}}(\vec{\chi}_{A},\vec{\chi}_{B})
  =
  \frac{d}{\sum_{k}|\chi_{k}^{(A)}\chi_{k}^{(B)}|^{2}}
  \sum_{k_{A}\neq k_{B}}|\chi_{k_{A}}^{(A)}\chi_{k_{B}}^{(B)}|^{2} \projector{e_{k_{A}},e_{k_{B}}}.
\end{equation}
The error operator $\widehat{\mathcal{E}}_{\mathbb{M}}$ characterizes the unexpected outputs for the target state $\ket{\psi}$ in the computation basis, which still contribute to the expectation value of the state verifier $\widehat{V}_{\mathbb{M}}$ in the subsequential measurements $\mathbb{M}$.
Employing the operators $\widehat{V}_{\psi}$ and $\widehat{\mathcal{I}}$, one can then estimate the lower and upper bounds on the $\ket{\psi}$-state fidelity $F_{\psi}$ according to Lemma \ref{lemma::SF_bounds}.
%\bigskip

In a laboratory, there will be a set of available measurement configurations $\mathbb{M}$.
However, taking all the available measurement configurations into the construction of the state verifier $\widehat{V}_{\mathbb{M}}$ does not always give us better bounds on the state fidelity.
Let $d=p_{1}^{n_{1}}...p_{k}^{n_{k}}$ be the prime number factorization of the local dimensionality $d$ with $p_{1}<...<p_{k}$.
One can show that the optimum bound on the state fidelity $F_{\psi}$ determined by Lemma \ref{lemma::SF_bounds} is achieved by the subsets $\widetilde{\mathbb{M}}$ of $\mathbb{M}$, which are constructed by selecting one element from each $p_{1}$-modulus equivalent subclass (quotient subset) $\mathbb{C}_{i}$ of $\mathbb{M}$.
Here, a~$p_{1}$-modulus subclass $\mathbb{C}_{i}$ of $\mathbb{M}$ is defined as
\begin{equation}
\label{eq::p_subcl}
  \mathbb{C}_{i}(\mathbb{M})
  :=
  \mathbb{M}\cap\left\{p_{1}k+i: k=0, ..., \frac{d}{p_{1}}-1\right\}
  \;\;\text{ with }\;\;
  i = 0, ..., p_{1}-1.
\end{equation}
From each nonempty subclass $\mathbb{C}_{i}$, one selects a measurement configuration to construct a subset $\widetilde{\mathbb{M}}$ of the available measurement configurations $\mathbb{M}$. The~set of all possible measurement configurations under this construction is
\begin{equation}
\label{eq::p1_msmnt}
  \bigotimes_{i=0, ..., p_{1}-1}
  \mathbb{C}_{i}(\mathbb{M})
  =
  \left\{
    \widetilde{\mathbb{M}}
    =
    \{j_{0},j_{1}, ..., j_{p_{1}-1}\}:
    j_{i}\in\mathbb{C}_{i}(\mathbb{M})
  \right\}.
\end{equation}
The cardinality of the subset of measurement configurations $\widetilde{\mathbb{M}}$ is equal to the number of nonempty $p_{1}$-modulus subclasses $\mathbb{C}_{i}$ of $\mathbb{M}$, which is denoted by $|\mathbb{M}_{/p_{1}}|$.
We can then assign a state verifier $\widehat{V}_{\widetilde{\mathbb{M}}}$ to each measurement configuration subset $\widetilde{\mathbb{M}}$ according to Equation \eqref{eq::psi_verifier_M} to determine a lower bound on $F_{\psi}$.
One can show that the optimum choice of the weights $\{u_{j}\}_{j\in\widetilde{\mathbb{M}}}$ for $\{\widehat{V}_{j}\}_{j\in\widetilde{\mathbb{M}}}$ in $\widehat{V}_{\widetilde{\mathbb{M}}}$ is the uniform weight $u_{j} = 1/|\mathbb{M}_{/p}|$, which takes the average of the state verifiers $\widehat{V}_{j}$ in $\widetilde{\mathbb{M}}$
\begin{equation}
  \widehat{V}_{\widetilde{\mathbb{M}}}(\vec{\chi}_{A},\vec{\chi}_{B})
  =
  \frac{1}{|\mathbb{M}_{/p_{1}}|}
  \sum_{j\in\widetilde{\mathbb{M}}}\widehat{V}_{j}(\vec{\chi}_{A},\vec{\chi}_{B}).
\end{equation}
As a result of Lemma \ref{lemma::SF_bounds}, one can estimate the lower and upper bounds on the $\ket{\psi}$-state fidelity as follows.
\begin{Theorem}[Lower and upper bounds on the state fidelity]
\label{theorem::fidelity_bounds}
  Let $\mathbb{M}\subseteq\{0,...,d-1\}$ be a set of measurement configurations associated with the local POVM measurements $\{\mathcal{M}[\widehat{\Omega}_{j}(\vec{\chi}_{A})]\otimes\mathcal{M}[\widehat{\Omega}_{-j}(\vec{\chi}_{B})]\}_{j\in\mathbb{M}}$, which are available in a laboratory.
  The state fidelity is then lower bounded by
\begin{equation}
  \label{eq::SF_LBound}
%    \braket{\widehat{V}_{\mathbb{M}}(\vec{\chi}_{A},\vec{\chi}_{B})} - \alpha_{\mathbb{M}}(\vec{u}) \braket{\widehat{\mathcal{E}}(\vec{\chi}_{A},\vec{\chi}_{B})}
    \max_{\widetilde{\mathbb{M}}\in \bigotimes_{i}\mathbb{C}_{i}(\mathbb{M})}
    \left(
      \Braket{\widehat{V}_{\widetilde{\mathbb{M}}}(\vec{\chi}_{A},\vec{\chi}_{B})}
    \right)
    -
    \frac{1}{|\mathbb{M}_{/p_{1}}|}
    \Braket{\widehat{\mathcal{E}}(\vec{\chi}_{A},\vec{\chi}_{B})}
    \le F_{\psi},
  \end{equation}
  and upper bounded by
\begin{equation}
  \label{eq::SF_UBound}
    F_{\psi} \le
    \min\left(
      \Braket{\widehat{V}_{e}},
      \min_{j\in\mathbb{M}}
      \braket{\widehat{V}_{j}(\vec{\chi}_{A},\vec{\chi}_{B})}
    \right).
  \end{equation}
  If $d$ is prime and $\mathbb{M}=\{0,...,d-1\}$, the~value of $F_{\psi}$ can be explicitly determined by
\begin{equation}
  \label{eq::SF_prime_d}
    F_{\psi}
    =
    \braket{\widehat{V}_{\mathbb{M}}(\vec{\chi}_{A},\vec{\chi}_{B})} - \frac{1}{d}\braket{\widehat{\mathcal{E}}(\vec{\chi}_{A},\vec{\chi}_{B})}.
  \end{equation}
  \par
%\noindent Proof: see \textbf{Methods}.
\end{Theorem}%

\begin{proof} see \textbf{Methods}.
\end{proof}
For a prime dimension $d$, the~lower and upper bounds on the state fidelity for the Bell-type state $\ket{\psi}$ in Equation \eqref{eq::SF_bounds} coincide with each other, which leads to an exact value of the state fidelity given in Equation \eqref{eq::SF_prime_d}.
One can therefore directly measure a state fidelity in $d+1$ measurement configurations.
In this case, the~method of state fidelity estimation in Equation \eqref{eq::SF_prime_d} is equivalent to the state fidelity derived in~\cite{BavarescoEtAlHuber2018-2MUBsCrtfyHghDmEnt}.
Since the state fidelity is exactly measured, the~choices of the coefficients $\vec{\chi}_{A,B}$ do not affect the final result\footnote{Theoretically, the~exact value of the $\ket{\psi}$-state fidelity of a testing state should not depend on the measurement configurations, if~one has large enough data of measurement outputs.}.
One can therefore choose $\vec{\chi}_{A,B}$ according to the feasibility of their corresponding measurement settings.
Note that the most simple measurement is usually the projective measurement with the uniform coefficient $\vec{\chi} = \{1/\sqrt{d}, ..., 1/\sqrt{d}\}$.
As a result of Equation \eqref{eq::chi_coeff_and_sCoeff}, the~preferable measurement settings in this case are then the projective measurements associated with $\vec{\chi}_{A} = \{1/\sqrt{d}, ..., 1/\sqrt{d}\}$ on one local system $A$ combined with the POVM measurements associated with $\vec{\chi}_{B} = \{s_{0}, ..., s_{d-1}\}$ on the other local system $B$.

If $d$ is non-prime or $\mathbb{M}\subsetneq\{0, ..., d-1\}$, there will be a gap between the lower and upper bounds on the state fidelity given in Equations \eqref{eq::SF_LBound} and \eqref{eq::SF_UBound}.
This gap can be reduced by carefully choosing proper coefficients $\vec{\chi}_{A,B}$ adapted to the measurement statistics in the computational basis before the implementation of remaining measurement configurations $\mathbb{M}(\vec{\chi}_{A},\vec{\chi}_{B})$.
Since the only information we have is the measurement statistics in the computational basis, we can not optimize $\vec{\chi}_{A,B}$ for the maximum expectation value of the state verifier $\widehat{V}_{M}$ that is evaluated in the upcoming measurements.
The optimization that we can carry out at this stage is to find the optimum $\vec{\chi}_{A,B}$ for the minimum expectation value of the error operator $\widehat{\mathcal{E}}(\vec{\chi}_{A},\vec{\chi}_{B})$ as follows:
\begin{gather}
\label{eq::opt_chiAB}
  (\vec{\chi}_{A}, \vec{\chi}_{B})
  = \argmin_{\vec{\chi}_{A}, \vec{\chi}_{B}}\braket{\widehat{\mathcal{E}}(\vec{\chi}_{A}, \vec{\chi}_{B})},
  \;\;\text{ subject to }\;\;
  s_{k} =
  \frac{\chi_{k}^{(A)}\chi_{k}^{(B)}}{\sqrt{\sum_{k}|\chi_{k}^{(A)}\chi_{k}^{(B)}|^{2}}}
  \text{ for all }k.
\end{gather}
The following conditions are sufficient for the minimum expectation value $\braket{\widehat{\mathcal{E}}}$
\begin{equation}
\label{eq::optimum_chi}
  \frac{|\chi_{k'}^{(A)}|^{2}}{|\chi_{k}^{(A)}|^{2}}
  =
  \sqrt{
    \frac{\prob_{e}(k,k')}{\prob_{e}(k', k)}
  }\frac{s_{k'}}{s_{k}}
  \;\;\text{ and }\;\;
  \frac{|\chi_{k'}^{(B)}|^{2}}{|\chi_{k}^{(B)}|^{2}}
  =
  \sqrt{
    \frac{\prob_{e}(k',k)}{\prob_{e}(k, k')}
  }\frac{s_{k'}}{s_{k}}
  \;\;\text{ for all } k,k'.
\end{equation}
However, these conditions can not be fulfilled for all $(k,k')$ in general.
For the special case when the measurement statistics is approximately symmetric under the exchange of the local systems, i.e.,~$\prob_{e}(k,k') \approx \prob_{e}(k',k)$, the~expectation value of the error operator is lower bounded by
\begin{equation}
  \braket{\widehat{\mathcal{E}}(\vec{\chi}_{A},\vec{\chi}_{B})}
  \ge
  \frac{d}{\sum_{k}s_{k}^{2}}
  \sum_{k_{A},k_{B}} s_{k_{A}}s_{k_{B}} \prob_{e}(k_{A},k_{B}),
%  ?(s_{k} , s_{k'})
%  d\sum_{k,k'}\sqrt{\prob_{e}(k,k)\prob_{e}(k',k')}\prob_{e}(k,k'),
\end{equation}
where the minimum is achieved by
\begin{equation}
\label{eq::opt_lambdaAB_symm}
  \chi^{(A)}_{k}
  =
  \chi^{(B)}_{k}
  =
  \sqrt{\frac{s_{k}}{\sum_{k}s_{k}}}.
  %\frac{\prob_{e}(k, k)^{1/4}}{\sqrt{\sum_{k}\sqrt{\prob_{e}(k, k)}}}.
\end{equation}
In practice, one may just want to estimate the state fidelity for the Bell-type state that is closest to the testing state, rather than a predefined one.
In this case, one can even adapt the Schmidt coefficients $s_{k}$ to the measurement probability $\prob_{e}(k,k)$ such that
\begin{equation}
\label{eq::sCoeff_prob_adapt}
  s_{k}
  =
  \sqrt{\frac{\prob_{e}(k,k)}{\sum_{k'}\prob_{e}(k',k')}}.
\end{equation}

\bigskip

As a whole, one can estimate a lower and an upper bound on the state fidelity for the Bell-type state that is closest to a testing state adaptively in the following~steps:
\begin{enumerate}
  \item One implements a measurement in the computational basis to obtain the statistics $\prob_{e}(k_{A},k_{B})$.
  \item Adapted to the measurement statistics $\{\prob_{e}(k_{A},k_{B})\}_{k_{A},k_{B}}$, one finds the optimum coefficients $\vec{\chi}_{A,B}$ for the minimum expectation value of the error operator $\braket{\widehat{\mathcal{E}}(\vec{\chi}_{A},\vec{\chi}_{B})}$ according to Equation~\eqref{eq::opt_chiAB}.
  \item Depending on the facilities of a laboratory, one implements a set of available local POVM measurements $\mathbb{M}(\vec{\chi}_{A},\vec{\chi}_{B})$ associated with the $\vec{\chi}_{A,B}$-modified Heisenberg--Weyl operators $\widehat{\Omega}_{j}(\vec{\chi}_{A})\otimes\widehat{\Omega}_{-j}(\vec{\chi}_{B})$ according to Equations \eqref{eq::ms_config} and \eqref{eq::HW_op_basis}.
      %and evaluate the corresponding correlation operators ;
  \item From the measurement statistic obtained in each measurement configuration $j\in\mathbb{M}(\vec{\chi}_{A},\vec{\chi}_{B})$, one evaluates the corresponding state verifier operator $\widehat{V}_{j}(\vec{\chi}_{A},\vec{\chi}_{B})$.
  \item Employing Theorem \ref{theorem::fidelity_bounds}, one estimates a lower and an upper bound on the state fidelity $F_{\psi}$.
\end{enumerate}

\subsection{Adaptive State Fidelity Estimation in Noisy Bell-Type State~Preparation}
\label{sec::SFE_noisy_BSt}

%\vspace{-6pt}

\begin{figure}
  \centering
%  \subfloat[]{
  \includegraphics[width=0.9\textwidth]{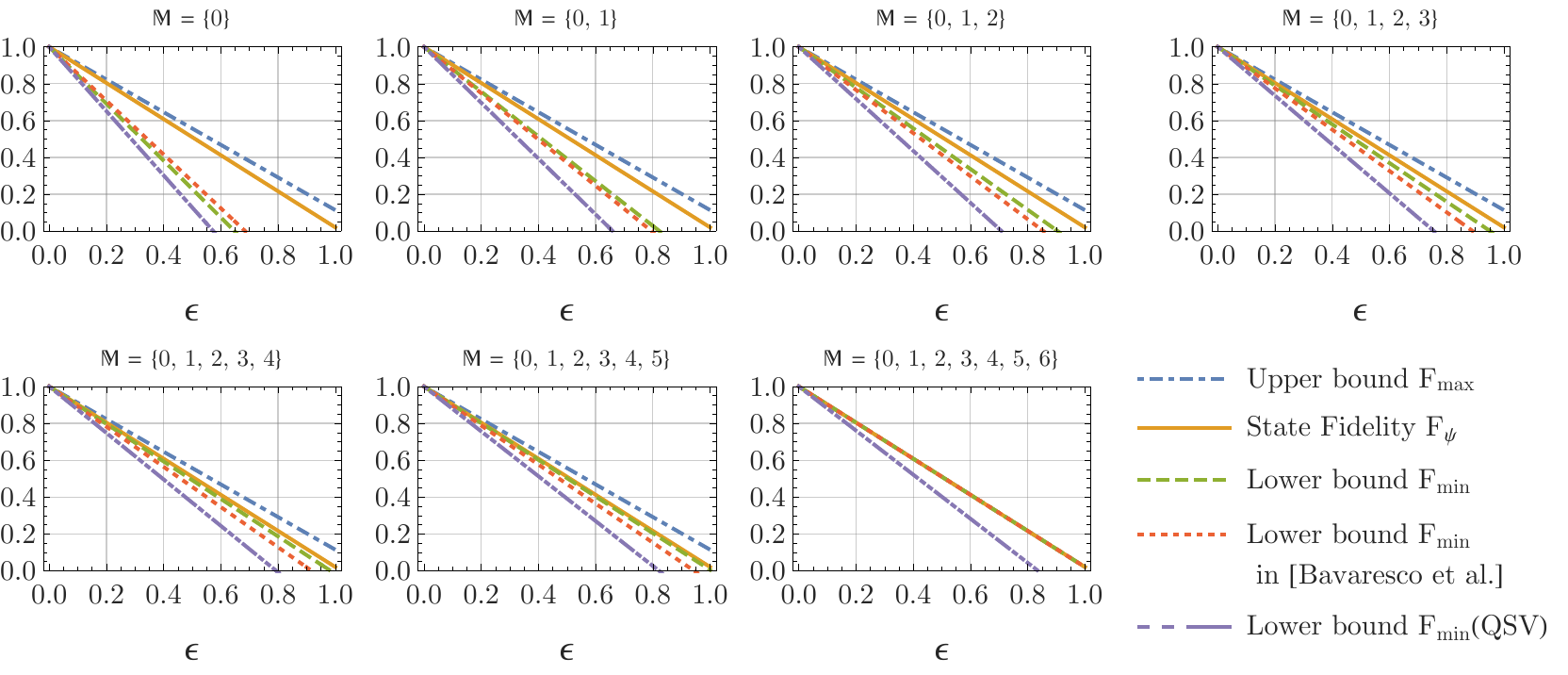}
%  }\\
%  \subfloat[]{
%  \includegraphics[width=\textwidth]{tildeSt_whiteNoise_d9}
%  }
  \caption{%
    Fidelity estimation of the noisy Bell-type state $\widehat{\rho}(\epsilon)$ in Equation \eqref{eq::white_noisy_Bell-type_st} for a $7\times 7$-dimensional Bell-type state $\ket{\psi}$ with the Schmidt coefficients $\{s_{k}\}_{k}=\{0.0845, 0.169, 0.254, 0.338, 0.423, 0.507, 0.592\}$ employing different measurement configurations.
    See the main text in Section~\ref{sec::SFE_noisy_BSt} for a detailed description.
%    These figures plot the state fidelity $F_{\psi}$ (orange solid), the fidelity upper bounds (blue dot-dashed) and lower bounds (green dashed) determined by Theorem \ref{theorem::fidelity_bounds}, the lower bounds derived in~\cite{BavarescoEtAlHuber2018-2MUBsCrtfyHghDmEnt} (red dotted line), and the lower bounds obtained by the non-adaptive method in Equation \eqref{eq::SF_bounds_nonadaptive} (violet dot-dot-dashed line).
%while the blue dot-dashed line and the green dashed line plots the fidelity upper bounds and lower bounds determined by Theorem \ref{theorem::fidelity_bounds}, respectively.
%The lower bounds are compared with the lower bounds derived in~\cite{BavarescoEtAlHuber2018-2MUBsCrtfyHghDmEnt} (red dotted line) and the ones obtained by the non-adaptive method in Equation \eqref{eq::SF_bounds_nonadaptive} (violet dot-dot-dashed line).
  }
  \label{fig::F_bounds_mconfigs_d7}
\end{figure}

In this section, we demonstrate the fidelity estimation method derived in Theorem \ref{theorem::fidelity_bounds} for Bell-type quantum states prepared under certain types of noises.
As an example, we first consider the white noises, which are symmetric under the exchange of two local systems.
In entanglement generation of a Bell-type state with the white noises, the~final state is described by
\begin{equation}
\label{eq::white_noisy_Bell-type_st}
  \widehat{\rho}(\epsilon)
  = (1-\epsilon)\projector{\psi} + \epsilon\frac{\widehat{\id}}{d^{2}},
%  \text{ with }
%  \ket{\psi} = \frac{1}{N_{\psi}}\sum_{k=0}^{d-1} (k+1)\ket{e_{k}, e_{k}}.
\end{equation}
where $\epsilon$ is the weight of the white noises.
The measurement statistics in the computational basis $\prob_{e}(k_{A},k_{B}) = \prob_{e}(k_{B},k_{A})$ is symmetric under the exchange of the local systems $A,B$.
One can therefore choose the measurement coefficients $\vec{\chi}_{A,B}$ as given in Equation \eqref{eq::opt_lambdaAB_symm}.
In this case, our approach employs the same measurement configurations as the ones employed in~\cite{BavarescoEtAlHuber2018-2MUBsCrtfyHghDmEnt}.
If one just exploits one measurement configuration added to the computational basis, the~lower bound derived in~\cite{BavarescoEtAlHuber2018-2MUBsCrtfyHghDmEnt} is tighter than the bound in Theorem \ref{theorem::fidelity_bounds}.
However, as~the number of measurement configurations in $\mathbb{M}$ increases, the~lower bound in Theorem \ref{theorem::fidelity_bounds} is improved faster, and~becomes better than the one derived in~\cite{BavarescoEtAlHuber2018-2MUBsCrtfyHghDmEnt}, which can be seen from the comparison between these two bounds in Figure~\ref{fig::F_bounds_mconfigs_d7} for a prime dimension $d=7$.

\begin{figure}
  \centering
%  \subfloat[]{\includegraphics[width=0.49\textwidth]{example-image-a}}
  \includegraphics[width=\textwidth]{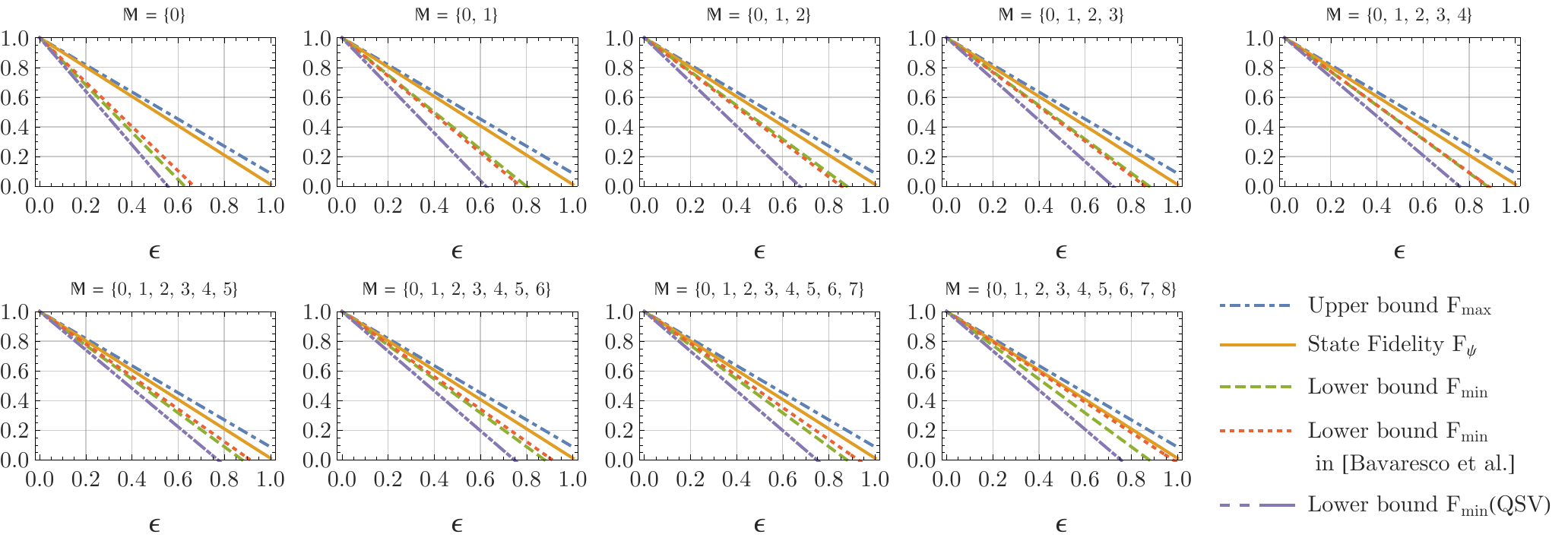}
  \caption{
    Fidelity estimation of the noisy Bell-type state $\widehat{\rho}(\epsilon)$ in Equation~\eqref{eq::white_noisy_Bell-type_st} for a $9\times 9$-dimensional Bell-type state $\ket{\psi}$ with the Schmidt coefficients $\{0.0592, 0.118, 0.178, 0.237, 0.296, 0.355, 0.415, 0.474, 0.533\}$ employing different measurement configurations.
    See the main text in Section~\ref{sec::SFE_noisy_BSt} for a detailed description.
%    The orange solid line plots the state fidelity $F_{\psi}$.
%    The blue dot-dashed line and the green dashed line plots the fidelity upper bounds and lower bounds determined by Theorem \ref{theorem::fidelity_bounds}, respectively.
%    The red dotted line plots the lower bound derived in~\cite{BavarescoEtAlHuber2018-2MUBsCrtfyHghDmEnt}, while the violet dot-dot-dashed line plots the lower bounds obtained by the non-adaptive method in Equation \eqref{eq::SF_bounds_nonadaptive}.
    %\textbf{The green dashed line plots the fidelity lower bounds derived in Theorem \ref{theorem::fidelity_bounds}, while the red dotted line plots the lower bound derived in~\cite{BavarescoEtAlHuber2018-2MUBsCrtfyHghDmEnt}, and the violet dot-dot-dashed line plots the lower bounds obtained by the non-adaptive method in Equation \eqref{eq::SF_bounds_nonadaptive}.
    %}
  }
  \label{fig::F_bounds_mconfigs_d9}
\end{figure}

In Figure~\ref{fig::F_bounds_mconfigs_d7}, we plot the state fidelity $F_{\psi}$ (orange solid) of a $7\times 7$-dimensional testing state $\widehat{\rho}(\epsilon)$, and~its corresponding upper (blue dot-dashed) and lower (green dashed) bounds determined by Theorem \ref{theorem::fidelity_bounds}.
These lower bounds are compared with the lower bounds derived in~\cite{BavarescoEtAlHuber2018-2MUBsCrtfyHghDmEnt} (red dotted) and the ones obtained by the nonadaptive method in Equation \eqref{eq::SF_bounds_nonadaptive} (violet dot-dot-dashed).
From this example, one can see that the lower bounds derived in Theorem \ref{theorem::fidelity_bounds} become tighter than the one in~\cite{BavarescoEtAlHuber2018-2MUBsCrtfyHghDmEnt}, if~one chooses more than one measurement configurations $\mathbb{M}\supseteq\{0,1\}$.
One can also see that both the adaptive methods in Theorem \ref{theorem::fidelity_bounds} and in~\cite{BavarescoEtAlHuber2018-2MUBsCrtfyHghDmEnt} can determine tighter lower bounds than the nonadaptive method in Equation \eqref{eq::SF_bounds_nonadaptive}.
%and implement a set of local POVM measurements $\mathcal{M}_{j}(\vec{\chi}_{A})\otimes\mathcal{M}_{-j}(\vec{\chi}_{B})$ with $j = 0, ..., d-1$ with $\chi^{(A)}_{k} = \chi^{(B)}_{k} = \sqrt{k+1}/\sqrt{d(d+1)/2}$.
%The lower bounds and upper bounds on the state fidelity for different measurement configurations are shown in Figure~\ref{fig::F_bounds_mconfigs} (a) and (b) for $d=7$ and $d=9$, respectively.
%
%Comparing the lower bound derived in~\cite{BavarescoEtAlHuber2018-2MUBsCrtfyHghDmEnt} with the one given in Theorem \ref{theorem::fidelity_bounds}, one can see that former one is tighter than the later for $\mathbb{M} = \{0\}$.
A limitation of the fidelity estimation in Theorem \ref{theorem::fidelity_bounds} is that, for a non-prime dimension, the~lower bounds are not necessarily tighter, if~the number of measurement configurations increases.
According to Theorem \ref{theorem::fidelity_bounds}, if~the available measurement configurations $\mathbb{M}\supset\{0,...,p_{1}-1\}$ have more than $p_{1}$ settings, then one should take the maximum of the lower bounds estimated by all subsets $\widetilde{\mathbb{M}}$ of $\mathbb{M}$, which has one element in each $p_{1}$-modulus subclass.
In this case, the~optimum lower bound obtained in Theorem \ref{theorem::fidelity_bounds} can be already saturated, when $\mathbb{M} = \{0,...,p_{1}-1\}$.
As one can observe in Figure~\ref{fig::F_bounds_mconfigs_d9} for $d=9$, the~optimum lower bounds on $F_{\psi}$ derived in Theorem \ref{theorem::fidelity_bounds} are already achieved by $\mathbb{M} = \{0, 1, 2\}$, while the lower bounds derived in~\cite{BavarescoEtAlHuber2018-2MUBsCrtfyHghDmEnt} are continuously improved, as~one includes more measurement configurations.
When one includes enough measurement configurations such that $\mathbb{M}\supseteq\{0,...,5\}$, the~method in~\cite{BavarescoEtAlHuber2018-2MUBsCrtfyHghDmEnt} can provide tighter lower bounds than the ones derived in Theorem \ref{theorem::fidelity_bounds}, while for the measurement configurations $\{0,1\}\subseteq\mathbb{M}\subseteq\{0, ..., 4\}$, the~method in Theorem \ref{theorem::fidelity_bounds} is still~better.

\begin{figure}[H]
  \centering
  \subfloat[]{\includegraphics[width=0.33\textwidth]{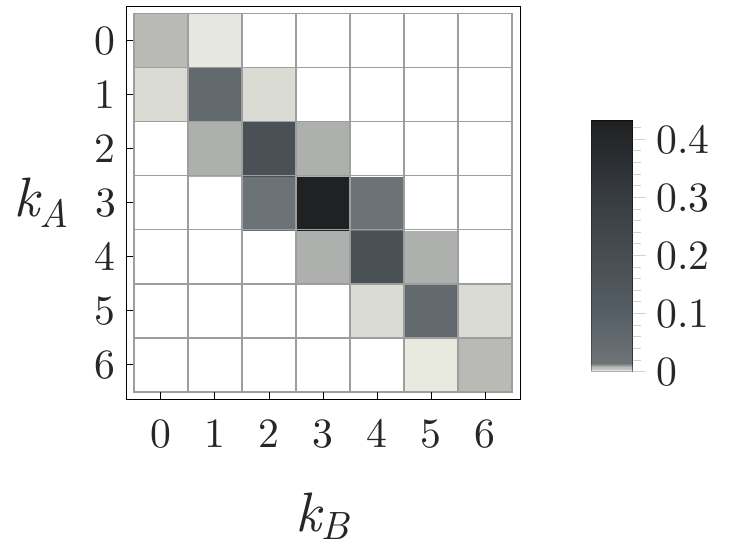}}
  \subfloat[]{\includegraphics[width=0.33\textwidth]{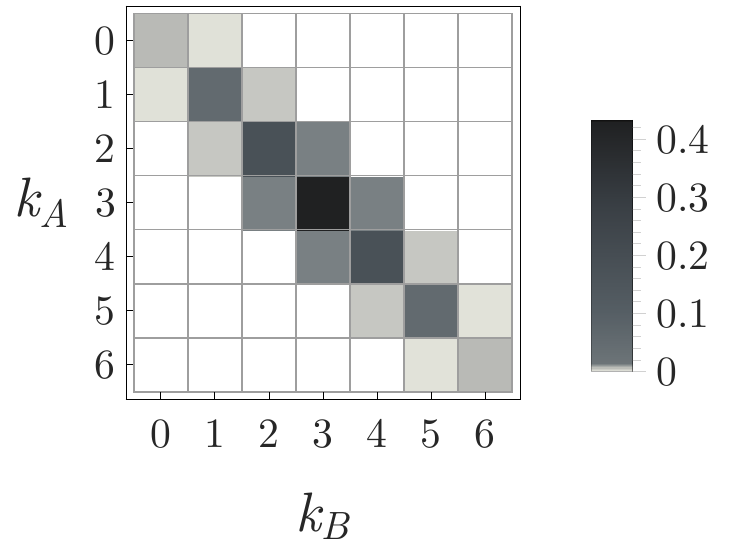}}
  \subfloat[]{\includegraphics[width=0.33\textwidth]{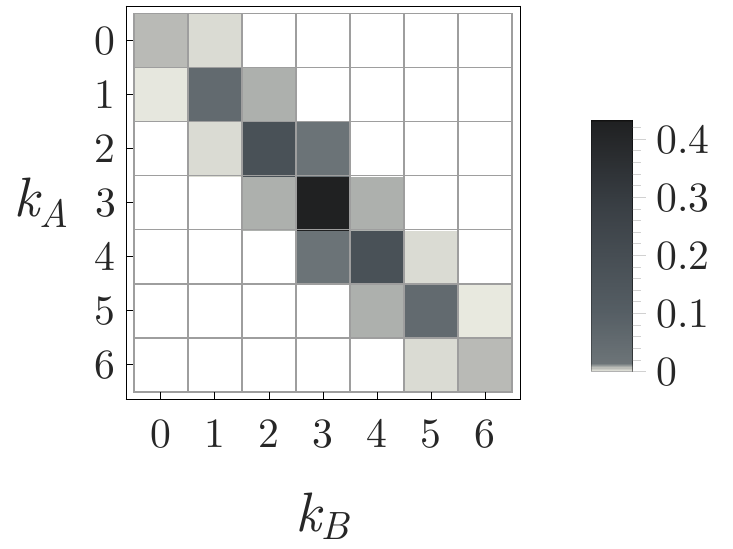}}
  \\
  \subfloat[]{\includegraphics[width=0.75\textwidth]{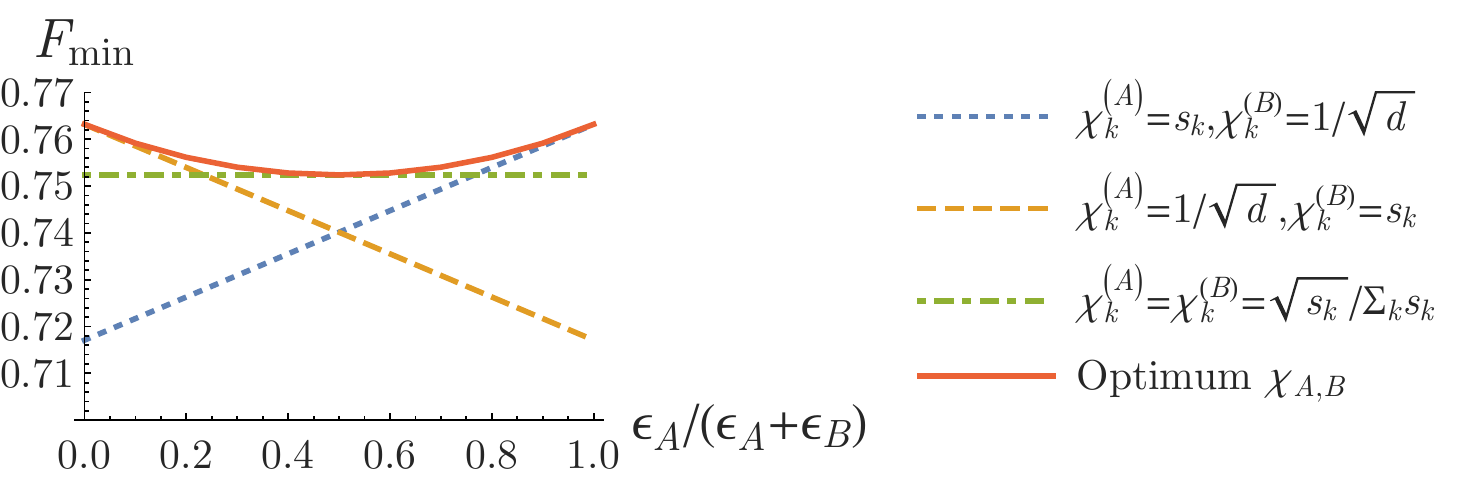}}
  \caption{
    Fidelity estimation for $\ket{\psi}$ with the Schmidt coefficients $\{s_{k}\}_{k} = \{0.086, 0.243, 0.446, 0.686, 0.446, 0.243, 0.086\}$  in a $(7\times 7)$-dimensional system under local crosstalk error model given in Equation \eqref{eq::cross_talk_noises}.
    The figures (\textbf{a}--\textbf{c}) show the measurement statistics of the state $\widehat{\rho}(\epsilon_{A},\epsilon_{B})$ in the computational basis with the local cross-talking errors $(\epsilon_{A},\epsilon_{B})$ of $(0.04,0)$, $(0.02,0.02)$, and~$(0,0.04)$, respectively.
    (\textbf{d}) The fidelity lower bounds estimated in Theorem \ref{theorem::fidelity_bounds} by choosing the measurement coefficients $(\vec{\chi}_{A}, \vec{\chi}_{B})$, which are determined for the one-side crosstalk in Equation \eqref{eq::asym_opt_chiAB} (blue dotted and orange dashed), the~symmetric crosstalk in Equation \eqref{eq::sym_opt_chiAB} (green dot-dashed), and~the general optimum in Equation \eqref{eq::opt_lambdaAB} (red solid), respectively.
  }
  \label{fig::lambda_compare}
\end{figure}

In general, the~noises in two separated local systems $A,B$ are not symmetric under the exchange of local systems.
In this case, Theorem \ref{theorem::fidelity_bounds} allows us to adapt the measurement coefficients $\vec{\chi}_{A,B}$ to the measurement statistics in the computational basis to refine the state fidelity estimation.
For example, in~linear optics networks~\cite{ReckEtAlBertani1994-ExpUniOP} which have path modes as their degree of freedom, one possible type of error is crosstalk between the computational basis states associated with neighboring paths.
If~the crosstalk error is small enough, such that the crosstalk between the computational-basis states $\ket{e_{k}}$ and $\ket{e_{k'}}$ associated with far neighboring paths $|k-k'|>1$ is negligible relative to the crosstalk between the closest neighboring paths $|k-k'|=1$, i.e.,~$\prob_{e}(k,k \pm \Delta k) \ll  \prob_{e}(k,k \pm 1)$ for $\Delta k>1$,
the expectation value $\braket{\widehat{\mathcal{E}}}$ can be approximately given by
\begin{equation}
  \braket{\widehat{\mathcal{E}}(\vec{\chi}_{A},\vec{\chi}_{B})}
  \approx
  \frac{d}{\sum_{k}|\chi_{k}^{(A)}\chi_{k}^{(B)}|^{2}}
  \sum_{k,k': |k-k'|=1} |\chi_{k}^{(A)}\chi_{k'}^{(B)}|^{2} \prob_{e}(k,k').
\end{equation}
In this case, the~optimum $\vec{\chi}_{A,B}$ determined by Equations \eqref{eq::optimum_chi} and \eqref{eq::sCoeff_prob_adapt} can be solved by
\begin{align}
\label{eq::opt_lambdaAB}
  \chi_{k}^{(A)}  & =
  \frac{1}{N_{A}}
  \left(
    \prob_{e}(k,k)
    \prod_{k'=0}^{k-1}\frac{\prob_{e}(k',k'+1)}{\prob_{e}(k'+1,k')}
  \right)^{1/4},
  \nonumber\\
  \chi_{k}^{(B)} & =
  \frac{1}{N_{B}}
  \left(
    \prob_{e}(k,k)
    \prod_{k'=0}^{k-1}\frac{\prob_{e}(k'+1,k')}{\prob_{e}(k',k'+1)}
  \right)^{1/4},
\end{align}
where $N_{A,B}$ are the normalization factors.
As an example, a~state produced in a Bell-type state generation under a simple model of local cross-talking noises $(\epsilon_{A},\epsilon_{B})$ can be described by
\begin{align}
\label{eq::cross_talk_noises}
  \rho_{\psi}(\epsilon_{A},\epsilon_{B})
  & =
  (1-2d(\epsilon_{A}+\epsilon_{B}))\projector{\psi}
  \nonumber\\
  & +
  \epsilon_{A}\sum_{k=0}^{d-1}
  \left(
    \ket{e_{k\oplus 1}^{(A)}}
    \braket{e_{k}^{(A)}|\psi}
    \braket{\psi|e_{k}^{(A)}}
    \bra{e_{k\oplus1}^{(A)}}
    +
    \ket{e_{k\ominus1}^{(A)}}
    \braket{e_{k}^{(A)}|\psi}
    \braket{\psi|e_{k}^{(A)}}
    \bra{e_{k\ominus1}^{(A)}}
  \right)
  \nonumber\\
  & +
  \epsilon_{B}\sum_{k=0}^{d-1}
  \left(
    \ket{e_{k\oplus1}^{(B)}}
    \braket{e_{k}^{(B)}|\psi}
    \braket{\psi|e_{k}^{(B)}}
    \bra{e_{k\oplus1}^{(B)}}
    +
    \ket{e_{k\ominus1}^{(B)}}
    \braket{e_{k}^{(B)}|\psi}
    \braket{\psi|e_{k}^{(B)}}
    \bra{e_{k\ominus1}^{(B)}}
  \right)
  .
\end{align}
Here, the~error coefficients $\epsilon_{A}$ and $\epsilon_{B}$ are the probability of a photon crossing to a closest neighboring path in the local system $A$ and $B$, respectively.
According to Equation \eqref{eq::opt_lambdaAB}, the~optimum $\vec{\chi}_{A, B}$ for one-side cross-talking errors are
\begin{equation}
\label{eq::asym_opt_chiAB}
  \left\{
    \begin{array}{ll}
      \chi_{k}^{(A)} = \sqrt{\frac{\prob_{e}(k,k)}{\sum_{k}\prob_{e}(k,k)}}, \chi_{k}^{(B)} = 1/\sqrt{d},
      & \hbox{for $\epsilon_{A}>0, \epsilon_{B}=0$;} \\
      \chi_{k}^{(A)} = 1/\sqrt{d}, \chi_{k}^{(B)} = \sqrt{\frac{\prob_{e}(k,k)}{\sum_{k}\prob_{e}(k,k)}},
      & \hbox{for $\epsilon_{A}=0, \epsilon_{B}>0$.}
    \end{array}
  \right.
\end{equation}
For symmetric cross-talking errors $\epsilon_{A} = \epsilon_{B}$, the~probability distribution $\prob_{e}(k,k')$ is symmetric under the exchange of $A$ and $B$, the~minimum of $\braket{\widehat{\mathcal{E}}}$ is then achieved by the measurement coefficients
\begin{equation}
\label{eq::sym_opt_chiAB}
  \chi_{k}^{(A)} = \chi_{k}^{(B)} = \sqrt{\frac{\sqrt{\prob_{e}(k,k)}}{\sum_{k}\sqrt{\prob_{e}(k,k)}}}.
\end{equation}
The computational-basis measurement statistics of the testing states $\widehat{\rho}(0.04,0)$ and $\widehat{\rho}(0,0.04)$ with one-side local crosstalk is asymmetric (see Figure~\ref{fig::lambda_compare}a,c), while it is symmetric for the testing state $\widehat{\rho}(0.02,0.02)$ with symmetric cross-talking errors (see Figure~\ref{fig::lambda_compare}b).
The lower bounds obtained by the different choices of measurement coefficients $\vec{\chi}_{A,B}$ given in Equations \eqref{eq::asym_opt_chiAB} and \eqref{eq::sym_opt_chiAB} are compared in Figure~\ref{fig::lambda_compare}d, where we fix the total cross-talking probability by $\epsilon_{A} + \epsilon_{B} = 0.04$ and calculate the fidelity lower bounds for different values of the ratio $\epsilon_{A}/(\epsilon_{A}+\epsilon_{B})$.
One can observe a $1.4\%$ improvement on the lower bound estimation, if~one chooses the optimum coefficients $\vec{\chi}_{A,B}$ in Equation \eqref{eq::opt_lambdaAB}, rather than the symmetric coefficients in Equation \eqref{eq::sym_opt_chiAB} for the one-side cross-talking errors $(\epsilon_{A} = 0.04, \epsilon_{B}=0)$ and $(\epsilon_{A} = 0, \epsilon_{B}=0.04)$.

\section{Discussion}

In this paper, we have employed state verifiers in Lemma \ref{lemma::Vj_contruction} to derive lower and upper bounds on state fidelity in Lemma \ref{lemma::SF_bounds}, which can be refined under the assistance of measurement statistics in the computational basis.
This method allows us to adapt the subsequential measurement configurations to measurement statistics in the computational basis to obtain tighter bounds, which are desirable for entanglement detection.
We have therefore employed this method to derive an adaptive approach of quantum state fidelity estimation for Bell-type bipartite entangled states in Theorem \ref{theorem::fidelity_bounds}.
This adaptive approach can determine lower and an upper bounds on the state fidelity which are tighter than the fidelity bounds obtained in QSV~\cite{ZhuHayashi2019-EffStVrfPRA}.
One has to note that QSFE and QSV have different problem settings.
We can not simply employ our adaptive method in QSV, since, in QSV, a~priori knowledge of a testing quantum system is not justified, and~the computational-basis measurement with a large enough number of outputs is inefficient.
To be precise, our method is good for the determination of tighter bounds on quantum state fidelity with the cost of some degree of inefficiency in the measurement process for obtaining a priori information in the computational~basis.

Another adaptive method of state fidelity estimation for Bell-type states is also derived in~\cite{BavarescoEtAlHuber2018-2MUBsCrtfyHghDmEnt}.
Their fidelity lower bound is tighter than the one derived in Theorem\ref{theorem::fidelity_bounds}, if~one implements only one measurement configuration added to the measurement in the computational basis.
However, our method can be tighter than their bound, if~one constructs more than one additional measurement configurations (see Figure~\ref{fig::F_bounds_mconfigs_d7}).
Note that our method can saturate its optimum, if~the set of measurement configurations $\mathbb{M}\subseteq\{0, ..., d-1\}$ is a $p_{1}$-modulus subclass of $\{0,...,d-1\}$, where $p_{1}$ is the smallest prime divisor of $d$.
For example, the~measurement configurations $\mathbb{M} = \{0, ..., p_{1}-1\}$ can already achieve the optimum of the QSFE employing Theorem \ref{theorem::fidelity_bounds} (see Figure~\ref{fig::F_bounds_mconfigs_d9}).
In this case, the~adaptive method in~\cite{BavarescoEtAlHuber2018-2MUBsCrtfyHghDmEnt} could provide tighter bounds again, if~one includes enough measurement configurations $\mathbb{M}\supset\{0, ..., p_{1}-1\}$ (e.g., $\mathbb{M}\supseteq\{0,...,5\}$ for QSFE of the $9\times 9$ quantum system demonstrated in Figure~\ref{fig::F_bounds_mconfigs_d9}).
However, to~get this benefit from the method in~\cite{BavarescoEtAlHuber2018-2MUBsCrtfyHghDmEnt} over our method in Theorem \ref{theorem::fidelity_bounds}, one will need to implement at least $(p_{1}+1)$ measurement settings added to the computational-basis~measurement.

One advantage of the adaptive method in this paper is that one can tailor the measurement configurations to asymmetric local noises by adjusting the coefficients $\vec{\chi}_{A,B}$ according to Equation~\eqref{eq::opt_chiAB}.
For the state preparation under a simple local cross-talking error model in linear optics network systems, where the probability of a photon in a path mode cross-talking with its neighboring paths is small enough, one can find the optimum local measurement coefficients $\vec{\chi}_{A,B}$ as given in Equation \eqref{eq::opt_lambdaAB}.
As shown in the example of Figure~\ref{fig::lambda_compare}d, the~optimum measurement coefficients $\vec{\chi}_{A,B}$ can improve the fidelity lower bound by about $1.4\%$ over the symmetric measurement~coefficients.

The approach in this paper only adapts to the measurement in the computational basis. It can be extended to a scheme of sequentially adaptive quantum state fidelity estimation analogous to the sequentially adaptive QST~\cite{KalevHen2015-FdltyOptQST}, in~which one constructs each subsequential measurement setting adapted to the measurement statistics obtained in all the prior measurement settings.
This adaptive scenario can be also extended to the QSV with a priori knowledge of measurement statistics in some particular bases, if~the a priori knowledge is justified by a trusted authorized~agent.

%%%%%%%%%%%%%%%%%%%%%%%%%%%%%%%%%%%%%%%%%%
\section{Methods}
\label{sec::methods}
In this section, we will prove our main results given in Lemmas \ref{lemma::Vj_contruction}, \ref{lemma::SF_bounds} and Theorem \ref{theorem::fidelity_bounds}.
First, we show the state verifier $\widehat{V}_{j}$ constructed in Lemma \ref{lemma::Vj_contruction} stabilizes the target state $\ket{\psi}$.
\begin{proof}[Proof of Lemma \ref{lemma::Vj_contruction}]
  After performance of the state verifier $\widehat{V}_{j}$ on the target state $\ket{\psi}$, one has
\begin{equation}
    \widehat{V}_{j}\ket{\psi}
    =
    \sum_{m_{A},m_{B}}
    \frac{v_{m_{A}m_{B}}}{d^{2}}
    \ket{E_{m_{A}}(j),E_{m_{B}}(j)}
    \braket{E_{m_{A}}(j),E_{m_{B}}(j)|\psi}.
  \end{equation}
  Since $\braket{e_{m_{A}},e_{m_{B}}|\widehat{T}_{A}^{\dagger}(j)\otimes\widehat{T}_{B}^{\dagger}(j)|\psi}
  =\braket{E_{m_{A}}(j),E_{m_{B}}(j)|\psi}$,
  with the construction given in Equation \eqref{eq::Vj_constr_vmAmB}, $\widehat{V}_{j}\ket{\psi}$, can be simplified to
\begin{equation}
    \widehat{V}_{j}\ket{\psi}
    =
    \sum_{m_{A},m_{B}}
    \ket{E_{m_{A}}(j),E_{m_{B}}(j)}
    \braket{e_{m_{A}},e_{m_{B}}|\widehat{T}_{A}^{-1}(j)\otimes\widehat{T}_{B}^{-1}(j)|\psi}.
  \end{equation}
  Since $\widehat{T}_{A}(j)\otimes\widehat{T}_{B}(j) = \sum_{m_{A},m_{B}}\ket{E_{m_{A}}(j),E_{m_{B}}(j)}\bra{e_{m_{A}},e_{m_{B}}}$ by definition, one can obtain the following eigenequation:
\begin{equation}
    \widehat{V}_{j}\ket{\psi}
    =
    \ket{\psi}.
  \end{equation}
  The operator $\widehat{V}_{j}$ constructed in Lemma \ref{lemma::Vj_contruction} is therefore a valid state verifier associated with the measurement $\mathcal{M}_{j}^{(A)}\otimes\mathcal{M}_{j}^{(B)}$.
\end{proof}

\bigskip

Second, we show the existence of a nontrivial decomposition $\ket{\widetilde{\phi}_{i}}_{i}$ for Lemma \ref{lemma::SF_bounds} and prove that the coefficients $\alpha$ and $\beta$ lead to the fidelity bounds given in \eqref{eq::SF_bounds}.
\begin{proof}[Proof of Lemma \ref{lemma::SF_bounds}]
  According to Equation \eqref{eq::psi_Phi_decomp}, the~operator $\widehat{V}_{\psi}^{\bot} - c \widehat{\mathcal{I}}^{\bot}$ can be decomposed as
\begin{equation}
  \label{eq::proof_lemma_1}
    \widehat{V}_{\psi}^{\bot} - c \widehat{\mathcal{I}}^{\bot}
    =\sum_{i}(\widetilde{\lambda}_{i} - c r_{i})\projector{\widetilde{\phi}_{i}}.
  \end{equation}
  For $c=\alpha:=\max_{i}(\widetilde{\lambda}_{i}/r_{i})$, the~operator $(c\widehat{\mathcal{I}}^{\bot}-\widehat{V}_{\psi}^{\bot})$ is positive semidefinite, while, for $c=\beta:=\min_{i}(\widetilde{\lambda}_{i}/r_{i})$, the~operator $(\widehat{V}_{\psi}^{\bot} - \beta\widehat{\mathcal{I}}^{\bot})$ is positive semidefinite.
  As a result of Equation \eqref{eq::V_psi_bot_bound}, the~two coefficients $\alpha$ and $\beta$ lead to the lower and upper bounds on the state fidelity given in Equation \eqref{eq::SF_bounds}.

  In the following steps, we show the existence of a nontrivial decomposition of $\widehat{V}_{\psi}^{\bot}$ and $\widehat{\mathcal{I}}^{\bot}$ for the fidelity estimation in Equation \eqref{eq::SF_bounds} by providing a protocol to find a decomposition $\{\ket{\widetilde{\phi}_{i}}\}_{i}$ and determine the corresponding coefficient $(\alpha,\beta)$.
  \begin{enumerate}
    \item One constructs a set of pure states $\Phi_{\mathbb{M}}$ for the decomposition of $\widehat{V}_{\mathbb{M}}^{\bot}$ through an extension of the $\widehat{V}_{j}$ eigenstates by local Pauli $\widehat{Z}$ operators
\begin{equation}
          \widehat{\Phi}_{\mathbb{M}} :=
          \bigcup_{j,k}\{\widehat{Z}^{m_{A}}\otimes\widehat{Z}^{m_{B}}\ket{\widetilde{\phi}_{j,k}}\}_{m_{A},m_{B}=0, ..., d-1}
          \;\;\text{ with }\;\;
          \widehat{Z} := \sum_{k}e^{\frac{\imI 2\pi}{d}k} \projector{e_{k}},
        \end{equation}
where $\{\ket{\widetilde{\phi}_{j,k}}\}_{k}$ are the eigenstates of $(\widehat{V}_{j}-\projector{\psi})$.~Employing the set~of~states~$\Phi_{\mathbb{M}}$, the~$\ket{\psi}$-orthogonal operator $\widehat{V}_{\psi}^{\bot}$ is then decomposed as
\begin{equation}
        \label{eq::V_decomp_Z_class}
          \widehat{V}_{\psi}^{\bot}
          =
          u_{e}\sum_{k}\projector{\widetilde{\phi}_{e,k}}
          +
          (1-u_{e}) \sum_{\ket{\varphi}\in\Phi_{\mathbb{M}}}\widetilde{\lambda}_{\varphi}\projector{\varphi},
        \end{equation}
where $\{\ket{\widetilde{\phi}_{e,k}}\}_{k}$ are the eigenstates of $V_{e}^{\bot}$ and $\widetilde{\lambda}_{\varphi}\ge0$ are non-negative.
    \item One constructs the operator $\widehat{\mathcal{E}}_{\mathbb{M}}$
        within the disjoint $\{\widehat{Z}_{A},\widehat{Z}_{B}\}$-equivalent subclasses $\{\widehat{\Phi}_{\mu}\}_{\mu}$ of $\Phi_{\mathbb{M}}$.
  %      Let $\Phi_{\mu}$ be the disjoint $\{\widehat{Z}\otimes\id, \id\otimes\widehat{Z}\}$-quotient subclasses $\{\widehat{\Phi}_{\mu}\}_{\mu}$ of $\Phi_{\mathbb{M}} = \cup_{\mu}\Phi_{\mu}$,
        Here, we say that two states $\ket{\varphi_{1}}$ and $\ket{\varphi_{2}}$ in the set $\Phi_{\mathbb{M}}$ are $\{\widehat{Z}_{A},\widehat{Z}_{B}\}$-equivalent, if~there exists $(m_{A},m_{B})$ such that $\ket{\varphi_{2}} = \widehat{Z}^{m_{A}}\otimes\widehat{Z}^{m_{B}}\ket{\varphi_{1}}$ up to a global phase.
        The set $\Phi_{\mathbb{M}}$ is then the union of the disjoint subclasses $\{\Phi_{\mu}\}_{\mu}$,
\begin{equation}
          \Phi_{\mathbb{M}} = \cup_{\mu}\Phi_{\mu}
          \;\;\text{ with }\;\;
          \Phi_{\mu} :=
          \{\ket{\varphi}\in\Phi_{\mathbb{M}}: \exists m_{A},m_{B},\theta \text{ such that }\ket{\varphi}=e^{\imI \theta}\widehat{Z}^{m_{A}}\otimes\widehat{Z}^{m_{B}}\ket{\varphi_{\mu}}
          \}.
        \end{equation}
        The sum of the projectors associated with the states in $\Phi_{\mu}$ is diagonal in the computational basis.
        One can then construct the operator $\widehat{\mathcal{I}}$ by assigning a positive weight $r_{\mu}>0$ to each subclass $\widehat{\Phi}_{\mu}$,
\begin{equation}
        \label{eq::Phi_construction_Z_class}
          \widehat{\mathcal{I}}
          =
          \widehat{V}_{e} + \widehat{\mathcal{E}}_{\mathbb{M}}
          \;\;\text{ with }\;\;
          \widehat{\mathcal{E}}_{\mathbb{M}}
          =
          \sum_{\mu}r_{\mu}\sum_{\ket{\varphi}\in\Phi_{\mu}}\projector{\varphi}.
        \end{equation}
        The operator $\widehat{\mathcal{I}}^{\bot} = \widehat{V}_{e}^{\bot}+\widehat{\mathcal{E}}_{\mathbb{M}}$ constructed in this way can then be decomposed as
\begin{equation}
        \label{eq::psi_Phi_decomp_Z_class}
          \widehat{\mathcal{I}}^{\bot}
          =
          \sum_{k}\projector{\widetilde{\phi}_{e,k}}
          +
          \sum_{\ket{\varphi}\in\Phi_{\mathbb{M}}}r_{\varphi}\projector{\varphi}
          \;\;\text{ with }\;\;
          r_{\varphi}\in\{r_{\mu}\}_{\mu}.
        \end{equation}
    \item As a result of the decompositions in Equations \eqref{eq::V_decomp_Z_class} and \eqref{eq::psi_Phi_decomp_Z_class}, the~coefficients $\alpha$ and $\beta$ are then determined by
\begin{equation}
          \alpha = \max\left(u_{e},(1-u_{e})\max_{\ket{\varphi_{l}}\in\Phi_{\mathbb{M}}}\frac{\widetilde{\lambda}_{l}}{r_{l}}\right)
          \;\;\text{ and }\;\;
          \beta = \min\left(u_{e},(1-u_{e})\min_{\ket{\varphi_{l}}\in\Phi_{\mathbb{M}}}\frac{\widetilde{\lambda}_{l}}{r_{l}}\right).
        \end{equation}
  \end{enumerate}
\end{proof}

\bigskip

For the proof of Theorem \ref{theorem::fidelity_bounds}, we need to find a proper decomposition $\{\ket{\widetilde{\phi}_{i}}\}_{i}$ of the state verifier given in Equation \eqref{eq::st_verifier_Btype}.
These state verifiers can be decomposed into a mixture of the generalized Bell-type states $\{\ket{\psi_{\mu\nu}(\vec{\chi}_{A}, \vec{\chi}_{B})}\}_{\mu,\nu}$ modified by the coefficients $\vec{\chi}_{A,B}$, which are defined as follows:
\begin{equation}
\label{eq::general_Bell_st}
  \ket{\psi_{\mu\nu}(\vec{\chi}_{A}, \vec{\chi}_{B})} :=
  \frac{1}{\sqrt{N(\mu)}}\sum_{k} w^{-\nu k} \chi_{k+\mu}^{(A)}\chi_{k}^{(B)}\ket{e_{k\oplus\mu},e_{k}}
  \;\;\text{ with }\;\;
  N(\mu;\vec{\chi}_{A},\vec{\chi}_{B}):=\sum_{k}|\chi_{k\oplus\mu}^{(A)}\chi_{k}^{(B)}|^{2}.
\end{equation}
Note that $\ket{\psi_{00}}$ is identical to the generalized Bell-type state $\ket{\psi}$ given in Equation \eqref{eq::Schmidt_decomp} %Please specify the formula number. We found that you deleted the equation.
if~the coefficient $\vec{\chi}_{A,B}$ are chosen according to Equation \eqref{eq::chi_coeff_and_sCoeff}.
Since the states $\ket{\psi_{\mu\nu}}$ are the eigenstates of the $(\vec{\chi}_{A},\vec{\chi}_{B})$-modified HW operators
\begin{equation}
  \widehat{\Omega}_{i,j}(\vec{\chi}_{A},\vec{\chi}_{B})\otimes\widehat{\Omega}_{i,-j}(\vec{\chi}_{A},\vec{\chi}_{B}) \ket{\psi_{\mu\nu}}
  =
  w^{\mu j + \nu i} \ket{\psi_{\mu\nu}(\vec{\chi}_{A},\vec{\chi}_{B})},
\end{equation}
the state verifier $\widehat{V}_{j}$ in Equation \eqref{eq::st_verifier_Btype} can be decomposed as
\begin{equation}
  \widehat{V}_{j}(\vec{\chi}_{A},\vec{\chi}_{B}) = \sum_{\mu j \oplus \nu = 0} \frac{N(\mu;\vec{\chi}_{A},\vec{\chi}_{B})}{N(0;\vec{\chi}_{A},\vec{\chi}_{B})}
  \projector{\psi_{\mu\nu}(\vec{\chi}_{A},\vec{\chi}_{B})},
\end{equation}
while the error operator $\widehat{\mathcal{E}}$ can be decomposed as
\begin{equation}
  \widehat{\mathcal{E}}(\vec{\chi}_{A},\vec{\chi}_{B})
  =
  \sum_{\mu\ge1,\nu}
  \frac{N(\mu;\vec{\chi}_{A},\vec{\chi}_{B})}{N(0;\vec{\chi}_{A},\vec{\chi}_{B})}
  \projector{\psi_{\mu\nu}(\vec{\chi}_{A},\vec{\chi}_{B})}.
\end{equation}
Employing these decompositions, one can prove Theorem \ref{theorem::fidelity_bounds} as follows.
\begin{proof}[Proof of Theorem 1]
  For the measurement configurations $\mathbb{M}(\vec{\chi}_{A}, \vec{\chi}_{B})\subseteq\{0, ..., d-1\}$ associated with the $(\vec{\chi}_{A},\vec{\chi}_{B})$-modified HW operators given in Equation \eqref{eq::ms_config}, one can construct a state verifier $\widehat{V}_{\psi}$ according to Equation \eqref{eq::psi_verifier_def} and decompose it into a mixture of the $(\vec{\chi}_{A},\vec{\chi}_{B})$-modified Bell-type states
\begin{equation}
    \widehat{V}_{\psi}
    =
    \projector{\psi}
    +
    u_{e}\sum_{\nu\ge1}\projector{\psi_{0,\nu}}
    +
    (1-u_{e})
    \sum_{\mu\ge1: \mu j\oplus \nu = 0}\frac{N(\mu)}{N(0)}u_{j}\projector{\psi_{\mu,\nu}}.
  \end{equation}
  Employing the same decomposition components $\{\ket{\psi_{\mu\nu}}\}_{\mu,\nu}$, the~operator $\widehat{\mathcal{I}}$ in Equation \eqref{eq::Phi_comp_basis} can be constructed by $\widehat{\mathcal{I}} = \widehat{V}_{e} + \widehat{\mathcal{E}}$.
%  \begin{equation}
%    \widehat{\mathcal{I}} = \widehat{V}_{e} + \widehat{\mathcal{E}}.
%  \end{equation}

  First, we derive the lower bound on the state fidelity as follows.
  According to Lemma \ref{lemma::SF_bounds}, the~coefficient $\alpha$ for the lower bound on $F_{\psi}$ is determined by
\begin{equation}
    \alpha(u_{e},u_{j\in\mathbb{M}}) =
    \max\left\{
      u_{e}, (1-u_{e})\widetilde{\alpha}
    \right\}
    \;\;\text{ with }\;\;
    \widetilde{\alpha}(u_{j\in\mathbb{M}}) := \max_{i}\left(\sum_{j\in\mathbb{C}_{i}(\mathbb{M})}u_{j}\right),
  \end{equation}
  where $\mathbb{C}_{i}(\mathbb{M})\in\mathbb{M}_{/p_{1}}$ are the $p_{1}$-modulus equivalent subclasses of the measurement configurations $\mathbb{M}$.
  Here, $p_{1}$ is the minimum prime-number divisor of the dimension $d$.
%  The coefficient $\alpha$ depends on the weights $u_{e}$ and $u_{j\in\mathbb{M}}$ of the state verifier components in $\widehat{V}_{\psi}$.
  The smaller the coefficient $\alpha$ is, the~larger the lower bound.
  The optimum lower bound is then obtained by the minimum value of $\alpha$, which is achieved by $\alpha = u_{e} = \widetilde{\alpha}/(1+\widetilde{\alpha})$.
  Insert this value of the coefficient $\alpha$ in Equation \eqref{eq::SF_bounds}, one~can obtain the lower bound
\begin{equation}
    \braket{\widehat{V}_{\mathbb{M}}(\vec{u}; \vec{\chi}_{A},\vec{\chi}_{B})}
    -
    \widetilde{\alpha}(u_{j\in\mathbb{M}})
    \braket{\widehat{\mathcal{E}}(\vec{\chi}_{A},\vec{\chi}_{B})}
    \le F_{\psi}
    \;\;\text{ with }\;\;
    \widehat{V}_{\mathbb{M}}(\vec{u}; \vec{\chi}_{A},\vec{\chi}_{B}) = \sum_{j\in\mathbb{M}}u_{j}\widehat{V}_{j}(\vec{\chi}_{A},\vec{\chi}_{B}).
  \end{equation}
  This bound can be improved by minimizing the coefficient $\widetilde{\alpha}$, which is equal to the number of nonempty $p_{1}$-modulus subclasses of $\mathbb{M}$, i.e.,~ $\min\tilde{\alpha} = |\mathbb{M}_{/p_{1}}|$.
  As a consequence, $F_{\psi}$ is lower bounded by
\begin{equation}
  \label{eq::thm1_proof_M_lbound}
    \braket{\widehat{V}_{\mathbb{M}}(\vec{\mu};\vec{\chi}_{A},\vec{\chi}_{B})}
    -
    \frac{1}{|\mathbb{M}_{/p_{1}}|}
    \braket{\widehat{\mathcal{E}}(\vec{\chi}_{A},\vec{\chi}_{B})}
    \le F_{\psi}.
  \end{equation}
  This lower bound is achieved by the weights $\sum_{j\in\mathbb{C}_{i}(\mathbb{M})}u_{j} = 1/|\mathbb{M}_{/p_{1}}|$, which are uniformly weighted over all $p_{1}$-modulus equivalent subclasses $\mathbb{C}_{i}(\mathbb{M})$.
  Indeed, this bound can be even improved by evaluating the lower bounds obtained with the measurement-configuration subsets $\widetilde{\mathbb{M}}(\subseteq\mathbb{M})$, which~have exactly one element in each $p_{1}$-modulus subclasses $\mathbb{C}_{i}(\mathbb{M})$ as given in Equation \eqref{eq::p1_msmnt}.
  With~each measurement configuration subset $\widetilde{\mathbb{M}}$, one determine a lower bound on $F_{\psi}$ employing the same formula given in Equation \eqref{eq::thm1_proof_M_lbound} with the measurement configuration weights $\{u_{j}=1/|\mathbb{M}_{/p_{1}}|\}_{j\in\widetilde{\mathbb{M}}}$.
  The~state verifier operator $\widehat{V}_{\widetilde{\mathbb{M}}}$ is then the average of the state verifiers $\widehat{V}_{j}$ associated with the measurement configurations in $\widetilde{\mathbb{M}}$,
\begin{equation}
    \widehat{V}_{\widetilde{\mathbb{M}}}
    =
    \frac{1}{|\mathbb{M}_{/p_{1}}|}
    \sum_{j\in\widetilde{\mathbb{M}}}\widehat{V}_{j}.
  \end{equation}
  As a result, one can estimate a lower bound on the state fidelity by
\begin{equation}
  \label{eq::thm1_proof_lbound}
    \max_{\widetilde{\mathbb{M}}\in \bigotimes_{i}\mathbb{C}_{i}(\mathbb{M})}
    \left(
      \Braket{\widehat{V}_{\widetilde{\mathbb{M}}}(\vec{\chi}_{A},\vec{\chi}_{B})}
    \right)
    -
    \frac{1}{|\mathbb{M}_{/p_{1}}|}
    \Braket{\widehat{\mathcal{E}}(\vec{\chi}_{A},\vec{\chi}_{B})}
    \le F_{\psi}.
  \end{equation}

  For the upper bound on $F_{\psi}$, one needs to determine the coefficient $\beta$ given in Equation \eqref{eq::alpha_beta} in Lemma \ref{lemma::SF_bounds}.
  If $d$ is non-prime or $\mathbb{M}\neq\{0, ..., d-1\}$, the~coefficient $\beta$ is always equal to zero.
  As a consequence, the~upper bound is given by the convex combination of the expectation values $\braket{\widehat{V}_{e}}$ and $\braket{\widehat{V}_{j\in\mathbb{M}}}$ weighted by $u_{e}$ and $u_{j\in\mathbb{M}}$, which means that $F_{\psi}$ is upper bounded by the minimum of $\braket{\widehat{V}_{e}}$ and $\braket{\widehat{V}_{j\in\mathbb{M}}}$,
\begin{equation}
    F_{\psi} \le
    \min\{\braket{\widehat{V}_{e}}, \min_{j\in\mathbb{M}}\braket{\widehat{V}_{j}(\vec{\chi}_{A},\vec{\chi}_{B})}
    \}.
  \end{equation}
  If $d$ is prime and $\mathbb{M} = \{0,...,d-1\}$, then the coefficient $\beta$ is given by
\begin{equation}
    \beta =
    \min\{u_{e}, \min_{j}(1-u_{e})u_{j}\}.
%    \left\{
%      \begin{array}{ll}
%        \min\{u_{e}, \min_{j}(1-u_{e})u_{j}\}, & \hbox{for $d$ prime and $\mathbb{M}=\{0, ..., d-1\}$;} \\
%        0, & \hbox{else.}
%      \end{array}
%    \right.
  \end{equation}
  The optimum choice of the weights $u_{e},u_{j\in\mathbb{M}}$ for the maximum $\beta$ is then $u_{e}=u_{j}= 1/(d+1)$, which~leads to the maximum $\beta = 1/(d+1)$.
  In this case, the~upper bound on $F_{\psi}$ determined in~Equation~\eqref{eq::SF_bounds}~is
\begin{equation}
  \label{eq::thm1_proof_ubound_prime_d}
    F_{\psi} \le
    \braket{\widehat{V}_{\mathbb{M}}(\vec{\chi}_{A},\vec{\chi}_{B})}
    -
    \frac{1}{d}\braket{\widehat{\mathcal{E}}(\vec{\chi}_{A},\vec{\chi}_{B})},
  \end{equation}
  where $\widehat{V}_{\mathbb{M}} = \sum_{j=0,...,d-1}\widehat{V}_{j}/(d+1)$ is the average of the state verifiers in the measurement configurations $\mathbb{M} = \{0,...,d-1\}$.
  Since the lower bound on $F_{\psi}$ given in Equation \eqref{eq::thm1_proof_lbound} coincides with the upper bound given in Equation \eqref{eq::thm1_proof_ubound_prime_d} for a prime $d$, the~state fidelity can be explicitly determined by the quantity given in Equation \eqref{eq::thm1_proof_ubound_prime_d}.
\end{proof}

%\bigskip
\vspace{+12pt}

%Materials and Methods should be described with sufficient details to allow others to replicate and build on published results. Please note that publication of your manuscript implicates that you must make all materials, data, computer code, and protocols associated with the publication available to readers. Please disclose at the submission stage any restrictions on the availability of materials or information. New methods and protocols should be described in detail while well-established methods can be briefly described and appropriately cited.
%
%Research manuscripts reporting large datasets that are deposited in a publicly available database should specify where the data have been deposited and provide the relevant accession numbers. If the accession numbers have not yet been obtained at the time of submission, please state that they will be provided during review. They must be provided prior to publication.
%
%Interventionary studies involving animals or humans, and other studies require ethical approval must list the authority that provided approval and the corresponding ethical approval code.

%%%%%%%%%%%%%%%%%%%%%%%%%%%%%%%%%%%%%%%%%%
%\section{Conclusions}
%
%This section is not mandatory, but can be added to the manuscript if the discussion is unusually long or complex.

%%%%%%%%%%%%%%%%%%%%%%%%%%%%%%%%%%%%%%%%%%
\vspace{-12pt}

%%%%%%%%%%%%%%%%%%%%%%%%%%%%%%%%%%%%%%%%%%
%% optional
%\supplementary{The following are available online at \linksupplementary{s1}, Figure S1: title, Table S1: title, Video S1: title.}


%%%%%%%%%%%%%%%%%%%%%%%%%%%%%%%%%%%%%%%%%%
%\authorcontributions{
%%For research articles with several authors, a short paragraph specifying their individual contributions must be provided. The following statements should be used ``Conceptualization, X.X. and Y.Y.; methodology, X.X.; software, X.X.; validation, X.X., Y.Y. and Z.Z.; formal analysis, X.X.; investigation, X.X.; resources, X.X.; data curation, X.X.; writing--original draft preparation, X.X.; writing--review and editing, X.X.; visualization, X.X.; supervision, X.X.; project administration, X.X.; funding acquisition, Y.Y. All authors have read and agreed to the published version of the manuscript.'', please turn to the  \href{http://img.mdpi.org/data/contributor-role-instruction.pdf}{CRediT taxonomy} for the term explanation. Authorship must be limited to those who have contributed substantially to the work reported.
%}

%%%%%%%%%%%%%%%%%%%%%%%%%%%%%%%%%%%%%%%%%%
\funding{This research is supported by JSPS International Research Fellowships (Standard), Grant No.  19F19817.

}
%\conflictsofinterest{{The authors declare no conflict of interest.}}

%%%%%%%%%%%%%%%%%%%%%%%%%%%%%%%%%%%%%%%%%%
%\acknowledgments{In this section you can acknowledge any support given which is not covered by the author contribution or funding sections. This may include administrative and technical support, or donations in kind (e.g., materials used for experiments).}

%%%%%%%%%%%%%%%%%%%%%%%%%%%%%%%%%%%%%%%%%%
%\conflictsofinterest{Declare conflicts of interest or state ``The authors declare no conflict of interest.'' Authors must identify and declare any personal circumstances or interest that may be perceived as inappropriately influencing the representation or interpretation of reported research results. Any role of the funders in the design of the study; in the collection, analyses or interpretation of data; in the writing of the manuscript, or in the decision to publish the results must be declared in this section. If there is no role, please state ``The funders had no role in the design of the study; in the collection, analyses, or interpretation of data; in the writing of the manuscript, or in the decision to publish the results''.}

%%%%%%%%%%%%%%%%%%%%%%%%%%%%%%%%%%%%%%%%%%
%% optional
\vspace{-6pt}
\abbreviations{The following abbreviations are used in this manuscript:\\

\noindent
\begin{tabular}{@{}ll}
QST & Quantum state tomography\\
QSV & Quantum state verification\\
QSFE & Quantum state fidelity estimation\\
HW operator & Heisenberg--Weyl operator\\
\end{tabular}}

%%%%%%%%%%%%%%%%%%%%%%%%%%%%%%%%%%%%%%%%%%
%% optional
\appendixtitles{no} %Leave argument "no" if all appendix headings stay EMPTY (then no dot is printed after "Appendix A"). If~the appendix sections contain a heading then change the argument to "yes".
\appendix

%\section{}
%\unskip
%\subsection{}
%The appendix is an optional section that can contain details and data supplemental to the main text. For example, explanations of experimental details that would disrupt the flow of the main text, but nonetheless remain crucial to understanding and reproducing the research shown; figures of replicates for experiments of which representative data are shown in the main text can be added here if brief, or as Supplementary data. Mathematical proofs of results not central to the paper can be added as an appendix.
%
%\section{}
%All appendix sections must be cited in the main text. In the appendixes, Figures, Tables, etc. should be labeled starting with `A', e.g.,~Figure A1, Figure A2, etc.

%%%%%%%%%%%%%%%%%%%%%%%%%%%%%%%%%%%%%%%%%%
\reftitle{References}

\end{document}